\documentclass[journal]{IEEEtran}
\usepackage{cite}
\usepackage{amsmath,amssymb,amsfonts}
\usepackage{algorithmic}
\usepackage{textcomp}
\usepackage{algorithm}
\usepackage{algorithmic}
\usepackage{mathrsfs}
\usepackage{booktabs}
\usepackage{graphicx}
\usepackage{subcaption}
\usepackage{longtable}
\usepackage{hyperref}
\usepackage{cleveref}
\usepackage{verbatim}
\usepackage{xcolor, soul}
\sethlcolor{yellow}

\begin{document}

\input epsf



\newtheorem{lemma}{Lemma}
\newtheorem{corol}{Corollary}
\newtheorem{theorem}{Theorem}
\newtheorem{proposition}{Proposition}
\newtheorem{definition}{Definition}

\title{6G Enabled Advanced Transportation Systems}

\author{Ruiqi Liu, Meng Hua, Ke Guan, Xiping Wang, 
Leyi Zhang, Tianqi Mao, \\ Di Zhang, Qingqing Wu and Abbas Jamalipour 
\thanks{R. Liu and L. Zhang are with the State Key Laboratory of Mobile Network and Mobile Multimedia Technology, Shenzhen 518055, China and the Wireless and Computing Research Institute, ZTE Corporation, Beijing 100029, China (e-mails: \{richie.leo, leyi.zhang\}@zte.com.cn).}
\thanks{M. Hua is with the Department of Electrical Engineering, City University of Hong Kong, Hong Kong  999077, China  (e-mail: menghua@cityu.edu.hk).}
\thanks{K. Guan and X. Wang are with the State Key Laboratory of Advanced Rail Autonomous Operation and the School of Electronic and Information Engineering, Beijing Jiaotong University, Beijing, 100044, China (e-mails: \{kguan, wangxiping\}@bjtu.edu.cn).}
\thanks{T. Mao is with the Advanced Research Institute of Multidisciplinary Science, Beijing Institute of Technology, Beijing 100081, China (e-mail: maotq18@163.com).}
\thanks{D. Zhang is with the School of Electrical and Information Engineering, Zhengzhou University, Zhengzhou 450001, China, and also with the School of Electrical Engineering, Korea University, Seoul 02841, Korea (e-mail: dr.di.zhang@ieee.org).}
\thanks{Q. Wu is with Department of Electronic Engineering, Shanghai Jiao Tong University, Shanghai 200240, China (e-mail: qingqingwu@sjtu.edu.cn).}
\thanks{A. Jamalipour is with School of Electrical and Information Engineering, University of Sydney, Camperdown NSW 2006, Australia (e-mail: a.jamalipour@ieee.org).}
\thanks{Corresponding author: Qingqing Wu.}}

\maketitle


\begin{abstract}
With the emergence of communication services with stringent requirements such as autonomous driving or on-flight Internet, the sixth-generation (6G) wireless network is envisaged to become an enabling technology for future transportation systems. In this paper, two ways of interactions between 6G networks and transportation are extensively investigated. On one hand, the new usage scenarios and capabilities of 6G over existing cellular networks are firstly highlighted. Then, its potential in seamless and ubiquitous connectivity across the heterogeneous space-air-ground transportation systems is demonstrated, where railways, airplanes, high-altitude platforms and satellites are investigated. On the other hand, we reveal that the introduction of 6G guarantees a more intelligent, efficient and secure transportation system. Specifically, technical analysis on how 6G can empower future transportation is provided, based on the latest research and standardization progresses in localization, integrated sensing and communications, and security. The technical challenges and insights for a road ahead are also summarized for possible inspirations on 6G enabled advanced transportation.
\end{abstract}


\section{Introduction}
Since the invention of wheels in the 4th millennium before the Common Era in Lower Mesopotamia, human society has been relying on many different types of transportation to meet the demand of moving around and transporting goods for daily uses and economic purposes. Transportation has been playing a vital role in modern lives and has profound impact to the economy and public welfare. The modes of transport commonly seen these days are land, air, water, and space transport, each provides a unique mobility and capacity. 

Another fundamental infrastructure that shapes our society drastically and creates a profound impact on daily life is the communication network. The history of human beings communicating through radio waves without a wire started in 1896 when Italian physicist Guglielmo Marconi invented the first successful wireless telegraph. Generation through generation, the functions of communication systems evolve from analog voice calls and instant text messages to Internet surfing and live streaming, even including more sophisticated functions such as timing and localization.
Recently, the 5th generation (5G) wireless networks have strengthened the traditional data service as well as supported many vertical industries by three major service categories, namely, the enhanced mobile broadband (eMBB), the ultra-reliable low latency communications (URLLC) and the massive machine type communications (mMTC). The initial phase of  5G deployment focuses on eMBB, which provides greater data rates complemented by moderate latency improvements. eMBB aims to satisfy most daily usage of 5G networks such as internet surfing, high-resolution streaming and downloading files. URLLC focuses on scenarios where reliability becomes a critical factor in communications, such as autonomous driving, where feedback and decisions must be transmitted and received with high reliability. mMTC can be seen as an enhancement towards full support of vertical industries by 5G networks, which aims to provide connections to a large number of devices in a certain area, such as sensors and robotics in smart factories. The current service categories of 5G can contribute to some regular transportation systems nowadays, such as buses and ferries, while transportation with a higher speed or higher altitude is sometimes out of reach by 5G.

As transportation systems grow larger and more complex, there is a natural trend for them to rely heavily on communication techniques. There has been literature on enhancing transportation systems with 5G or previous generations of wireless networks. One literature points out that inefficiencies in transportation systems could cause enormous losses of time, a decrease in the level of safety for both vehicles and pedestrians, high pollution, degradation of quality of life, and waste of energy \cite{5430544}. 
To battle the inefficiencies, an intelligent transportation system (ITS) is proposed to collect input from the sensors and devices on board and to produce helpful directives to drivers as well as to the transportation infrastructure to improve the efficiency and security level. 
The conventional ITS collects data from sources like inductive loop detectors and pneumatic tubes and processes data mainly based on historical and human experience, which leads to a relatively high implementation cost and limited performance. The development of communication technology, advanced data collection, data analysis, and data transmission ways enable a revolution from technology-driven ITS to data-driven ITS. 
Data collected from video cameras, the global positioning system (GPS) and  sensors ensure the amount of training samples of ITS to complete detection, recognition, and tracking of the traffic-related objects \cite{5959985}. 
The usage of artificial intelligence (AI) techniques in data processing allows ITS to deal with traffic problems like incident detection and traffic forecasting \cite{sumalee2018smarter}.
By exchanging information among different vehicles and infrastructures, the cognitive management functionality can be achieved not only inside the vehicle but also in the overall transportation infrastructure, helping drivers make decisions as well as enhancing the efficiency and safety of the transportation system \cite{dimitrakopoulos2010systems}.
However, the large amount of heterogeneous data brings difficulties to achieve a high data rate as well as a low latency. 
With the advent of 5G, novel technologies like device-to-device communication, multiple-input multiple-output (MIMO), millimeter Wave (mmWave) communication and small cells are proposed to address the above challenges, though future transportation systems would demand capabilities beyond 5G. 

\begin{figure}[t]
\centering
\includegraphics[width=0.95\linewidth]{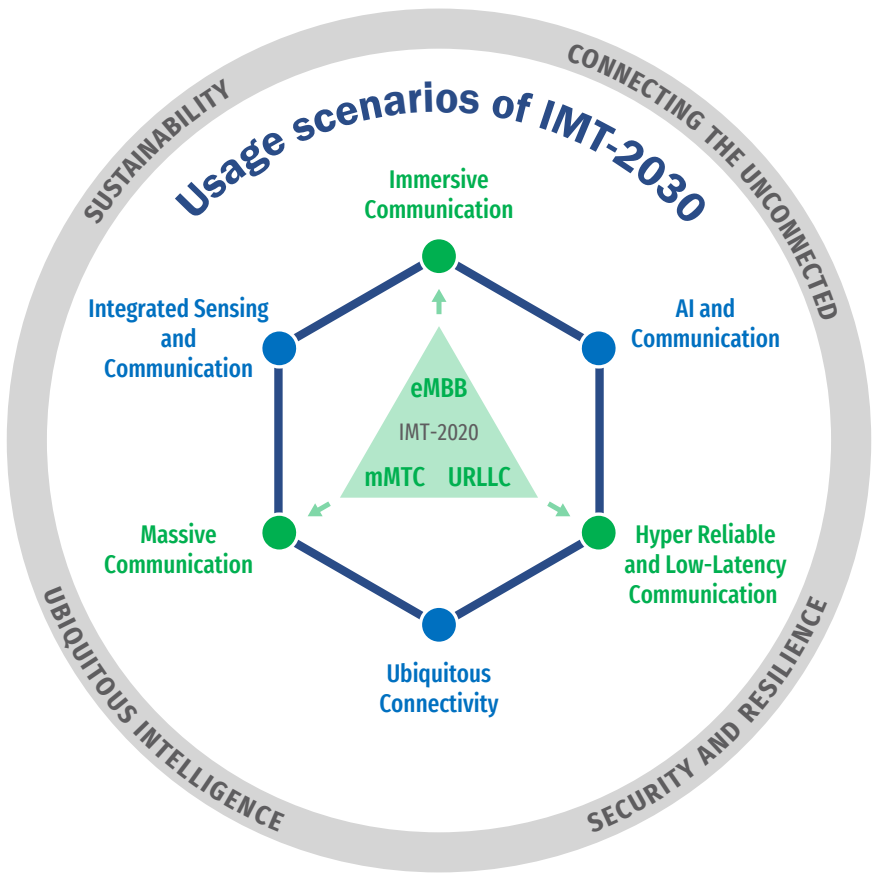}
\caption{Illustration of the evolution from the three usage scenarios of 5G to the six new usage scenarios of 6G.}
\label{fig:6G}
\end{figure}

Moving towards the 6th generation (6G), it is a broad consensus that wireless networks will continue to evolve on the three categories and provide higher data rate, connection density, and reliability, and at the same time reduce latency and power consumption. As depicted in Fig. \ref{fig:6G}, it is envisaged that the three basic usage scenarios of 5G will inherit to 6G and be enhanced, to meet more stringent requirements \cite{liu_2023_6G}. What's more, to better support vertical industries and their diverse needs, 6G will also emphasize three new scenarios, namely, integrated sensing and communication (ISAC), {AI and communication} along with ubiquitous connectivity. All of these usage scenarios, along with the corresponding representative capabilities, will surely open a new era of communication and boost vertical industries as well, such as transportation.

\begin{table}[t]
    \caption{\label{Tab:6G}Summary of key capability items of 6G compared to that of 5G as endorsed in the ITU recommendation
 \cite{10298069}}
    \centering
    \begin{tabular}{|p{0.35\columnwidth}|p{0.16\columnwidth}|p{0.3\columnwidth}|}
    \hline
        \bf{Capability items} & \bf{5G values} & \bf{6G exemplary values}  \\ \hline
        Peak data rate & 20 Gbps & 50, 100, 200 Gbps  \\ \hline
        User experienced data rate & 100 Mbps & 300, 500 Mbps  \\ \hline
        Spectrum efficiency & 3 $\times$ 4G & $(1.5 $ – $3) \times$ 5G  \\ \hline
        Area traffic capacity & 10 Mbit/s/m\textsuperscript{2} & 30, 50 Mbit/s/m\textsuperscript{2}  \\ \hline
        Connection density & $10^6$ devices/km\textsuperscript{2} & $10^6 - 10^8$ devices/km\textsuperscript{2}  \\ \hline
        Mobility & 500 km/h & 500 - 1000 km/h  \\ \hline
        Latency & 1 ms & 1 – 0.1 ms  \\ \hline
        Reliability & 1-10\textsuperscript{-5} & (1-10\textsuperscript{-5}) – (1-10\textsuperscript{-7})  \\ \hline
        Energy efficiency & 100 $\times$ 4G & No value \\ \hline
        Security and resilience & Qualitative & Qualitative  \\          \hline
        Positioning accuracy & - & 1-10 cm  \\ \hline
        Coverage & - & To be specified  \\          \hline
        Sensing-related capabilities & - & To be specified  \\ \hline
        Applicable AI-related capabilities & - & Qualitative  \\ \hline
    \end{tabular}
\end{table}

On top of serving more usage scenarios, 6G is envisaged to provide capabilities in multiple dimensions that exceed the 5G limit. To date, the vision and framework recommendation for 6G is stabilized {by} the ITU-R \cite{10298069}, with the key capability items summarized in Table \ref{Tab:6G}. Compared to the values in 5G, it can be seen that for data rate, connection density, and efficiency, 6G intends to enable capabilities around 5 times larger than its previous generation. Stronger support for maximal mobility is also a target, making it possible to connect devices on high-speed trains or even airplanes using terrestrial cellular networks. 6G also emphasizes capabilities related to security, resilience, sensing, positioning, and AI, opening up new possibilities for smarter and {securer} transportation systems. Note that the exemplary values given are just possible research targets, and higher values are also possible for future studies. Moreover, these capability items will be evaluated under different scenarios and are not meant to be satisfied at the same time.

6G is still being studied and discussed dynamically in 
{both} the academia and the industry, while some early results emerge to indicate that some candidate technologies might prevail in their time to support the six usage scenarios as well as the representative capabilities. In this paper, the latest research findings on some key candidate technologies for 6G are reported and demonstrated to show how 6G is able to provide connectivity to all forms of transportation, even in remote areas or under extreme conditions, and how 6G will contribute to {greener, smarter, securer and more robust transportation systems}.

{Since 6G and ITS are both emerging and dynamic topics, there exists little survey on the interplay between the two systems. This paper intends to provide a comprehensive survey as well as some insights for this area.}
The rest of the paper is structured as follows. How 6G will provide ubiquitous connectivity to all forms of transportation is elaborated in detail in Section \ref{Sec_2}. On top of that, Section \ref{Sec_3} introduces some key candidate technologies {for} 6G that can make transportation systems more intelligent, robust, and secure. Finally, Section \ref{Sec_conclusion} will conclude this paper.


\section{Ubiquitous connectivity for all transportation} \label{Sec_2}
As the number and type of connected devices grow rapidly, connectivity is demanded in all categories of transportation with large bandwidth, low latency, and high reliability. 
Currently, as 5G is being rolled-out throughout the world, modern time networks can already satisfy a part of the demand for connectivity on various kinds of transportation. It is no longer an imagination to watch anime with high resolution on trains, nor does texting friends and families on an airplane sound luxurious. However, there are still some transportation scenarios where connectivity is scarce or the bandwidth is extremely limited. In the coming future, it is envisaged that the 6G network will be able to provide ubiquitous connectivity to passengers or devices on all types of transportation in almost every corner of the world, including some of the most remote and unpopulated areas on this planet. Thanks to the novel designs and advanced techniques from physical layers to higher layers, 6G will strive to connect everyone and everything on the trip, including transportation forms elaborated below that challenge every aspect of a communication system.

\subsection{Railway transportation}
\begin{figure*}[t]
\centering
\includegraphics[width=0.6\linewidth]{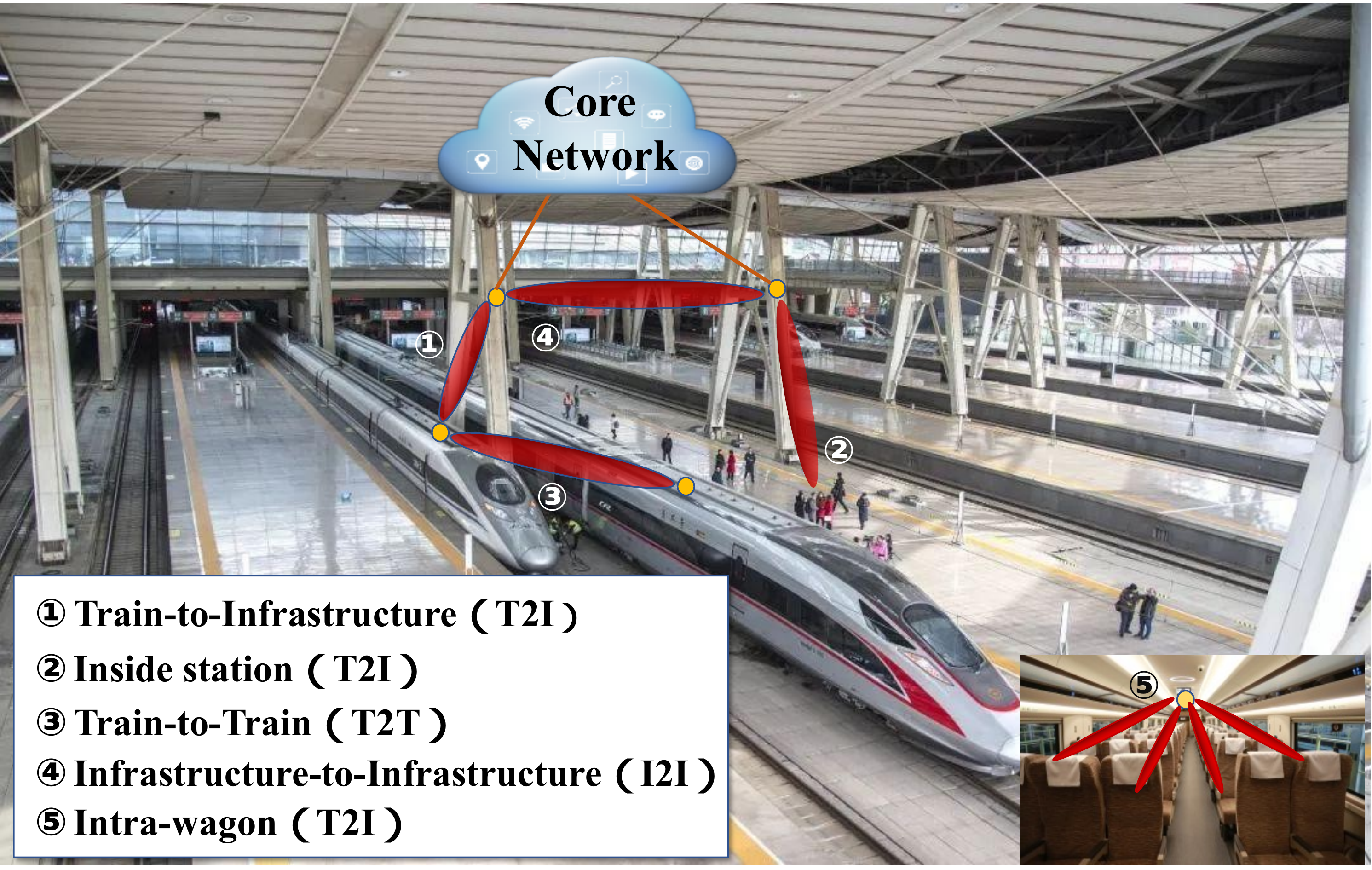}
\caption{The view of five smart rail mobility scenarios.}
\label{fig:srm}
\end{figure*}

Empowered by 6G, railway transportation is expected to evolve into the "smart rail mobility" era where infrastructure, trains, passengers, and goods will be fully interconnected \cite{KG_TVT_2021}. The present vision entails an upgrade in railway communications, which would expand their capacity beyond just managing essential signaling applications. The enhanced communication capabilities would support a range of  applications demanding high data rates, such as high-definition video streaming for passengers, dispatching high-definition video in real-time, and multimedia journey information \cite{KG_TVT_2017}. For smart rail mobility, as shown in Fig. \ref{fig:srm}, the high-data-rate transmission will be realized in five typical scenarios, namely, the train-to-infrastructure (T2I), inside station (IS), train-to-train (T2T), infrastructure-to-infrastructure (I2I), and intra-wagon (IW). 
To support a high peak data rate and user-experienced data rate in all the above scenarios, the required bandwidth ranges from several hundred MHz to several GHz. The need for bandwidth will be more significant in the T2T and IW scenarios, with dozens of GHz needed for each. The biggest challenge lies in the T2I scenario because a very high-data-rate, low latency, and uninterrupted service must be realized. As a result, to establish communication of up to 100 Gbps, we need a bandwidth of dozens of GHz or even higher \cite{KG_MITS_2021}. The considerable bandwidth and high-data-rate demands create a compelling incentive to leverage the terahertz (THz) and sub-THz band, which lies above 100 GHz and offers a vastly abundant spectrum \cite{THz_survey}. As sub-THz and THz bands are still new for the industry to explore, we will elaborate on these new frequency ranges a bit more and show how to effectively overcome some challenges faced by exploiting such high frequencies.
We will detail the channel conditions for all five smart rail mobility scenarios at the 300 GHz band as an example, introducing ultra-wideband (UWB) channel sounding, ray tracing (RT), channel characterization, and discussing how channel characteristics can influence the design and deployment of future railway communication systems.

\subsubsection{Channel characterization}
To serve the different communication scenarios, 6G will use the combination of all available bands, from low bands to sub-THz. Radio waves of lower frequency are ideal to provide wide coverage in open areas since the propagation attenuation is much lower. Thus, for trains passing through open terrains, low-band and mid-band can be used to provide ubiquitous connectivity, especially for services without the need for a high data rate.

\begin{figure*}[h]
\centering
\includegraphics[width=0.9\linewidth]{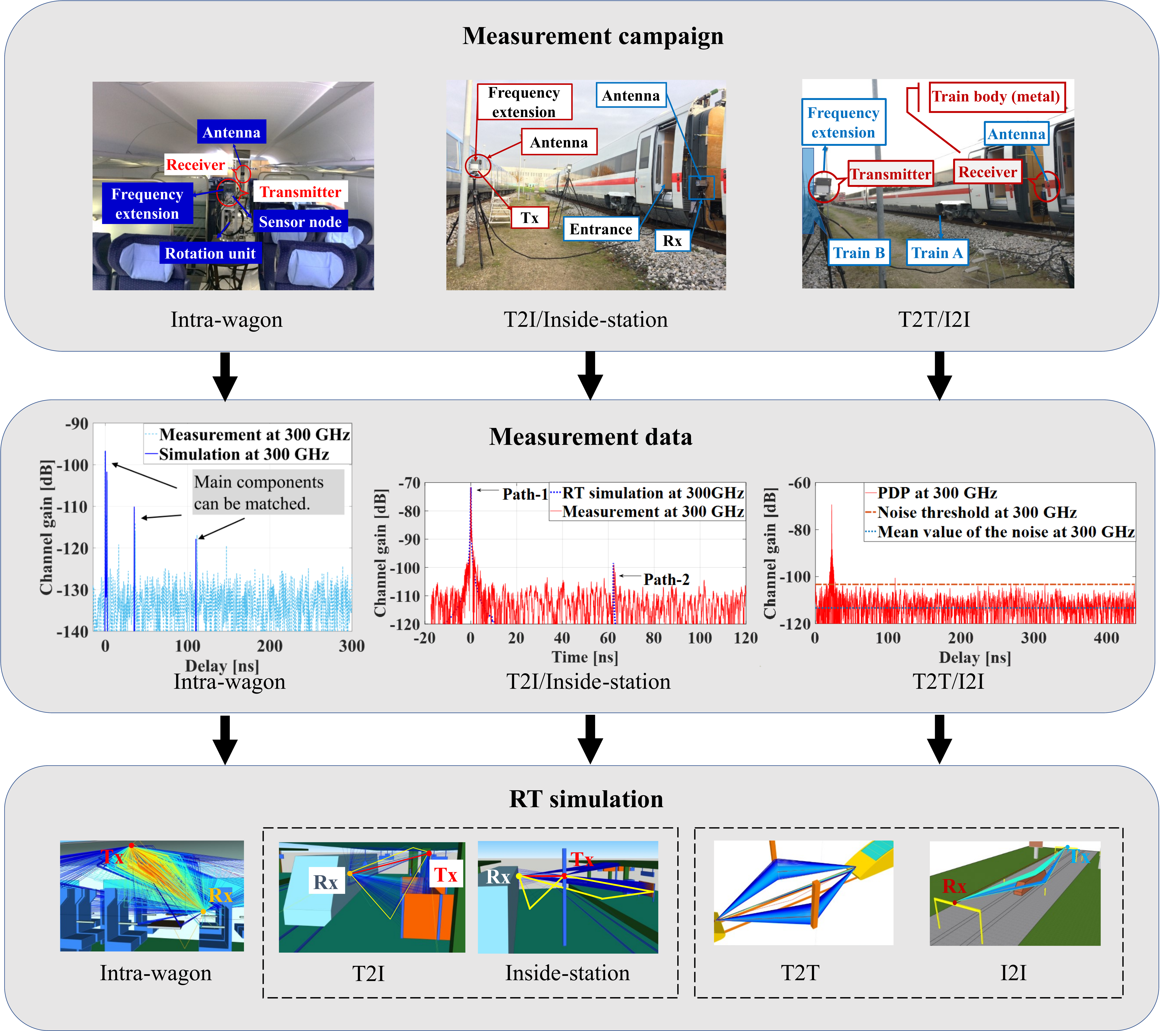}
\caption{An illustration of measurement campaign, measurement data, and RT simulations.}
\label{RTwork}
\end{figure*}

For services that desire a high peak data rate, sub-THz and THz bands will be beneficial.

\begin{table}[t]
\caption{\label{tab:300GHz}Channel measurement results at 300 GHz}
\centering
\begin{tabular}{|l|c|c|c|c|}
\hline
\multicolumn{1}{|c|}{\begin{tabular}[c]{@{}c@{}} \bf{Measurement}\\ \bf{Set Index}\end{tabular}} &
  \begin{tabular}[c]{@{}c@{}}\bf{Rician} \\ \bf{K-factor}\end{tabular} &
  \begin{tabular}[c]{@{}c@{}}\bf{RMS} \\ \bf{delay spread}\end{tabular} &
  \bf{ASA} &
  \bf{ASD} \\ \hline
1 (IW)        & 8.39 dB & 11.74 s & 36.68$^\circ$ & 18.75$^\circ$ \\ \hline
2 (T2I/IS) & 3.52 dB & 8.92 s & -      & -      \\ \hline
3 (T2T/I2I)            & 3.60 dB & 9.23 s  & -      & -      \\ \hline
\end{tabular}
\end{table}

All values of the Rician K-factor, root mean squared (RMS) delay spread, and azimuth angular spread of arrival/departure (ASA/ASD) extracted from the measurements are summarized in Table \ref{tab:300GHz}. The Rician K-factor at 300 GHz is 3.52 dB, and the RMS delay spread at 300 GHz is 8.92 ns, just 50\%  of the mean value of the RMS delay spread measured in a similar scenario at 60 GHz (16 ns). These results suggest that the multi-path propagation of the THz channel in the T2I/IS scenario is considerably sparser than that at lower frequencies. However, the channel still exhibits limited degrees of freedom, underscoring the need for measurement-validated real-time simulations to extend the preliminary channel characterization. More comprehensive channel characteristics for the five typical scenarios at the 300 GHz band are summarized in a previous study \cite{KG_TVT_2021}.




\subsubsection{Fast and adaptive beamforming}
MIMO has become a standard operation in 5G and will no doubt still be a critical component of 6G. For MIMO communication systems that wish to support high mobility scenarios such as high-speed trains, fast and adaptive beamforming is {desperately} desired because {UEs may move rapidly along with the high-speed trains.}
{Fast and adaptive beamforming requires angular domain information to rapidly adjust the antenna array weights in order to enhance the received signal strength at the locations of UEs and maximize the coverage of BS, thus optimize the performance of MIMO systems.} According to existing simulation results \cite{KG_TVT_2021}, angular spreads depend greatly on environments, and as a consequence, the values of these parameters vary significantly in different scenarios. In the IS and IW scenarios, the angular spreads in the azimuth direction are larger (16.9$^{\circ}$ - 39.8$^{\circ}$) than those in the elevation direction (2.0$^{\circ}$ - 6.3$^{\circ}$). This implies that in these two scenarios, most of the multi-paths are present in the azimuth plane, which can potentially provide more diversity to the MIMO system and require adaptive beamforming of a larger angular domain. However, for the T2T channel, the mean values of elevation angular spread of arrival/departure (ESA/ESD) are larger (32.4$^{\circ}$ and 31.6$^{\circ}$, respectively) compared to the mean values of ASA and ASD, which are 2.2$^{\circ}$ and 11.2$^{\circ}$, respectively. This relation implies a crucial characteristic of the T2T channel, where the multi-paths in the vertical planes (reflected or scattered from the tracks and the front of trains) dominate more than those in the azimuth plane. 
Thus, the fast and adaptive beamforming algorithms in 6G shall be designed with enough flexibility to adapt to different scenarios with unique propagation conditions.
  
\subsubsection{Mechanisms against the Doppler effect}
With the combination of the high mobility of the railway and potentially higher frequencies of wireless communication, the Doppler effect will be extremely severe. For instance,  for T2I and T2T scenarios at the 300 GHz band, the mean values of mean Doppler shifts can achieve 72.32 kHz and 8.25 kHz, respectively \cite{KG_TVT_2021}. This would exceed the capability of Doppler shift calibration in the current 5G system and thus, novel mechanisms against the Doppler effect must be designed.
Possible directions to investigate include a tailored design of sub-carrier spacing that is larger than the maximum of RMS Doppler spread in the order to dozens of kHz, novel transmission protocols with the aid of reconfigurable intelligent surfaces (RIS), and so on. 
The RIS is a large planar array comprising a large number of passive elements, each of which is able to independently adjust the impinging signal's amplitude and/or phase in real-time, achieving both passive beamforming gain and planar aperture gain \cite{wu2019intelligent,9679804,ris_field_test}. 
For instance, by deploying real-time tune-able RISs \cite{BoAi} in railway scenarios, the impact of the Doppler shift on the fast-fluctuated received signal power can be effectively reduced. Moreover, one favorable feature of the railway is that the location, direction, and velocity information {of the train} can be obtained and therefore, the Doppler shift can be known as prior information. This makes it easy to track and compensate for the Doppler effects by designing efficient algorithms of the adjustable {phase shifts of RIS elements} for high-speed trains {(HSTs)} \cite{JiayiZhang}. Apart from mechanism methods, data-driven paradigms can also help. For instance, taking advantage of the fact that an HST travels the same path repeatedly, based on the reference signal received power (RSRP) values measured by the mobile receiver on the HSTs at all times, machine learning can be {used} to develop fast and accurate Doppler shift estimators with low complexity \cite{Taehyung}.

\subsubsection{Proper modulation and channel coding}
Modulation is the process of modifying a carrier signal by varying one or more properties of the carrier signal. Channel coding adds error-correcting codes as redundancy to the transmitted signal to enable error correction at the Rx. The choices of modulation {schemes} and {coding rates} depend on the characteristics of the channel, such as the noise level and the presence of multi-path fading. {The} Rician K-factor measures the dispersion of the signal in a wireless communication system caused by multi-path fading. A lower Rician K-factor generally indicates greater dispersion, leading to increased multi-path interference, higher signal attenuation and reduced signal-to-noise ratio {(SNR)}. Accurate knowledge of the Rician K-factor is crucial for selecting the appropriate modulation and coding schemes. According to some recent results \cite{KG_TVT_2021}, the Rician K-factor is with a mean value larger than 5~dB in most LoS conditions of all five {smart rail mobility} scenarios, except for two cases in the IS scenario. These two cases are C-NT and C-T \cite{KG_TTHT_2019}, where C-NT represents the case where the Tx is deployed on the catenary mast between two tracks and there is no other train on the adjacent track. C-T refers to the case where the Tx is deployed on the catenary mast between two tracks and there is another train on the adjacent track. This implies that much stronger multi-paths can be received when the Tx is placed on the catenary mast. Additionally, unlike the T2T scenario, where the mean Rician K-factor is quite large (12.97 dB), the mean Rician K-factor in the LoS area of the IW scenario ranges from 4.27 dB to 6.90 dB, which is still smaller than that at lower frequencies. This implies that the multi-paths inside a wagon are still strong even at 300 GHz. As a result, in these cases where Rician K-factor is relatively lower, lower-order modulation schemes and lower coding rates are suggested in communication system design. {Furthermore}, apart from orthogonal frequency division multiplexing (OFDM), novel modulation schemes like orthogonal time frequency space (OTFS) modulation can be promising candidates to deal with the multipath channel with strong time-frequency selectivity in HSTs.  
In OTFS, the resource is multiplexed in the delay-Doppler (DD) domain in which the CIRs can be considered as time-invariant. Based on the realistic channel measurements, the OTFS is able to realize lower complexity of channel estimation, higher diversity, and higher reliability for HST communications when the speed is over 200 km/h \cite{Yiyan1}. When considering the requirement of massive internet of things (mIoT) with high mobility, a transceiver can be jointly designed with OTFS and tandem spreading multiple access (TSMA). Simulation results in an existing literature illustrate that both high user connectivity and transmission reliability in HST mIoT can be achieved by OTFS-TSMA \cite{Yiyan2}. 
As to potential new coding schemes for 6G, recent studies show that novel coding designs can introduce enhancements into several typical deployment scenarios, such as polarization adjusted convolutional codes with rate matching in hyper reliable and low-latency communication (HRLLC) scenarios \cite{10233770} and an amplitude phase shift keying (APSK) based scheme for immersive communication \cite{9824198}. These new designs will benefit transportation systems with different types of mobility.



\subsection{Air to ground communication}
Air-to-ground (ATG) communication has a long history dated back to the 1900s \cite{667666}. This particular type of communication used to be conducted based on the use of analogue voice signals on very high frequency (VHF) bands, mainly serving the air traffic control. This early version of ATG communication enables the air crew to report their position and status to ground controllers and for the controller to give appropriate and timely instructions to the air crew.

Nowadays, with the ever increasing demand of mobile data everywhere, there has been an emerging use case to provide connectivity not only to the crew but also the passengers on board. This demand is particularly popular on long haul flights where passengers may need to stay connected with families, colleagues or customers for important matters. The non-terrestrial network (NTN) is a plausible technical solution for this use case as described in subsection \ref{Sec_NTN}, thanks to its capability to provide connectivity to almost every location around the globe. However, due to the significant latency and low data rate of NTN, such networks are usually only capable of receiving and sending emails or chat messages, instead of video conferences, live streaming and gaming. 

\begin{figure}[ht]
\centering
\includegraphics[width=0.9\linewidth]{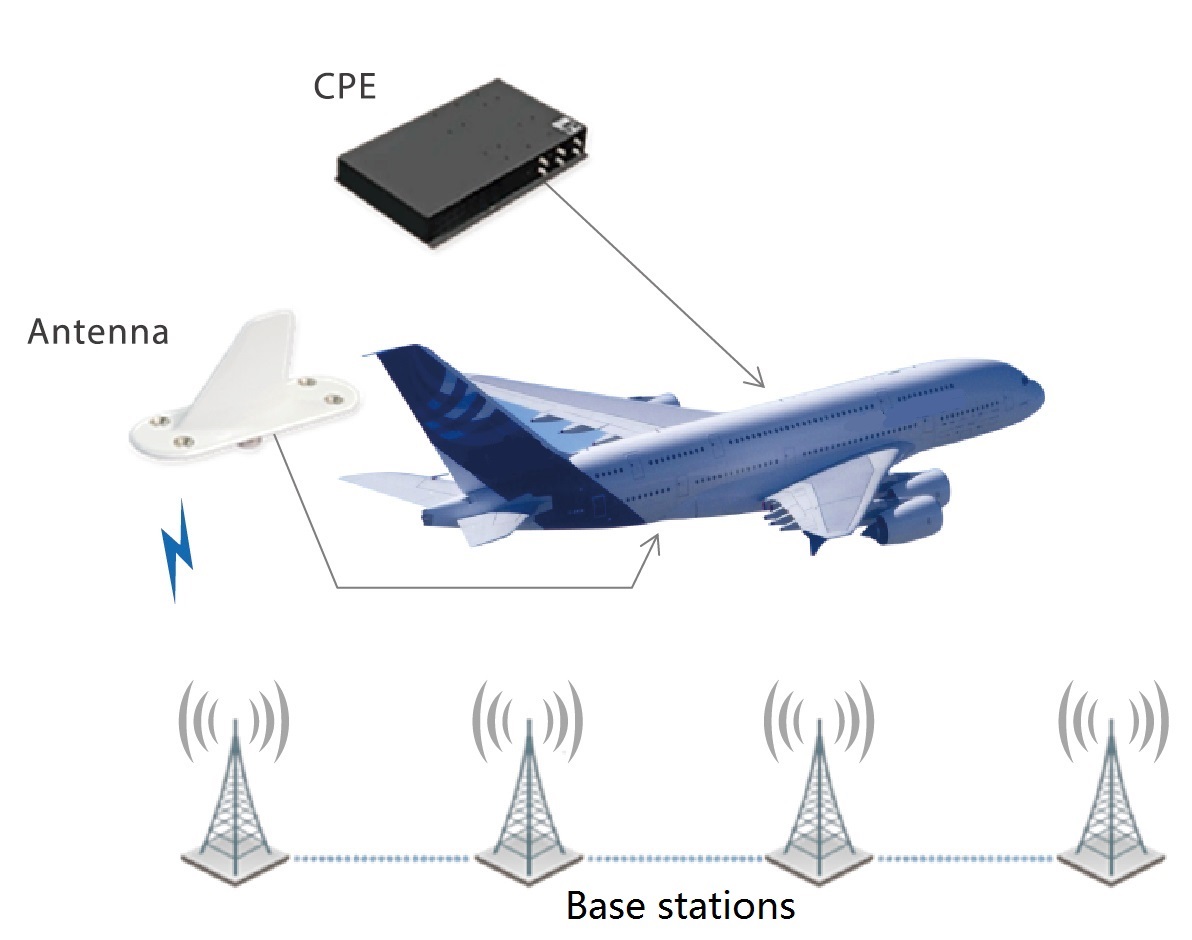}
\caption{Illustration of the architecture of a cellular ATG network.}
\label{fig:ATG}
\end{figure}

To provide connectivity to the crew and passengers on board an aircraft with satisfactory data rate and latency, cellular ATG networks have been proposed as a candidate solution. The idea is to serve the UEs on board a flight using BSs of the ground cellular networks such as 5G or 6G. 
Using mature cellular communication technologies, cellular ATG systems can serve high mobility scenarios with satisfactory quality of service (QoS). Special BSs that can cover the sky are established on the ground to enable high-bandwidth, high-throughput with reasonable cost of deployment, upgrade and maintenance. Such a system has the ability to provide airline passengers with on-board entertainment, on-board office and customized services. It has a wide range of industry application prospects, such as aviation medical rescue, flight operation and intelligent and digital aviation administration.
An illustration of a typical cellular ATG system is provided in Fig. \ref{fig:ATG}.
In the downlink, signals are transmitted from the BSs of cellular networks towards the airborne antenna, which is usually installed at the bottom of an airplane. The airborne antenna is connected to a customer premise equipment (CPE), which processes and converts the signal received from cellular networks to WiFi signals. Finally, the cabin users are served by the WiFi network for data and mobile services.


Doppler shift is a challenging issue in ATG systems. Due to the long distance between the aircraft and the BSs as well as ultra high mobility, novel technologies are required to guarantee user experience in the air. Apart from the novel Doppler compensation methods, some emerging technologies such as enhanced cell coverage, precise beam tracking and customized high-performance antenna, will be needed in the future.


\subsubsection{Enhanced cell coverage}
The radius or a cell in 5G is approximately several hundred meters, determined by the frequency band used. 6G will likely to exploit higher frequency bands, resulting in a similar cell radius. On the other hand, due to the high mobility of the aircraft, using an ordinary cell radius will cause frequent handover performed by the CPE and thus, significantly affect the performance and user experience of the cellular ATG system. 
Therefore, cellular ATG systems demand an enhanced cell coverage in order to reduce the frequency of handovers and ensure stable connections. 
To increase the cell coverage, there are some mature technologies in wireless communication and signal processing that can be beneficial, including enhanced frame structure design, physical random access channel (PRACH) transmission timing adjustment, uplink and downlink hybrid automatic repeat request (HARQ) and interference suppression. It is reported and demonstrated that some 5G based cellular ATG networks can have a maximum cell radius of 300 km \cite{ZTE_ATG}, and it's reasonable to expect more advanced system performance supported by 6G.

\subsubsection{Precise beam tracking}
When the aircraft is cruising, the signals transmitted from the BSs propagate mainly in a direct LoS path to the antenna on the airplane. Given the long distance between the ground BS and the aircraft, beamforming technique is mandatory to compensate the significant propagation attenuation introduced by the large distance. In 5G, beam training and tracking are done in a periodic way through transmission and reception of reference signals between the BS and the UE. For cellular ATG networks, to ensure the robustness of data reception by cabin users, it is proposed to adopt an adaptive beam adjustment scheme which can select the optimal beam width at cell center or cell edge \cite{9734362}. Due to the fact that the cruise path of an civil airplane is usually pre-determined and doesn't vary dramatically, a machine learning based predictive beam tracking can also be used instead of periodic beam training \cite{9410605}. Such machine learning based methods can learn the relationship between the historical positions and velocities and appropriate beam-steering angles, to facilitate a more accurate beam tracking with less overhead.

\subsection{Complementary coverage for transportation in remote areas} \label{Sec_NTN}
Transportation can take place in remote and unpopulated areas, such as ships sailing in oceans or visitors crossing the deserts. These areas are almost not served by terrestrial networks, and how to meet the demand for data on these kinds of transportation is also a mission for 6G.

\begin{figure*}[ht]
\centering
\includegraphics[width=0.8\linewidth]{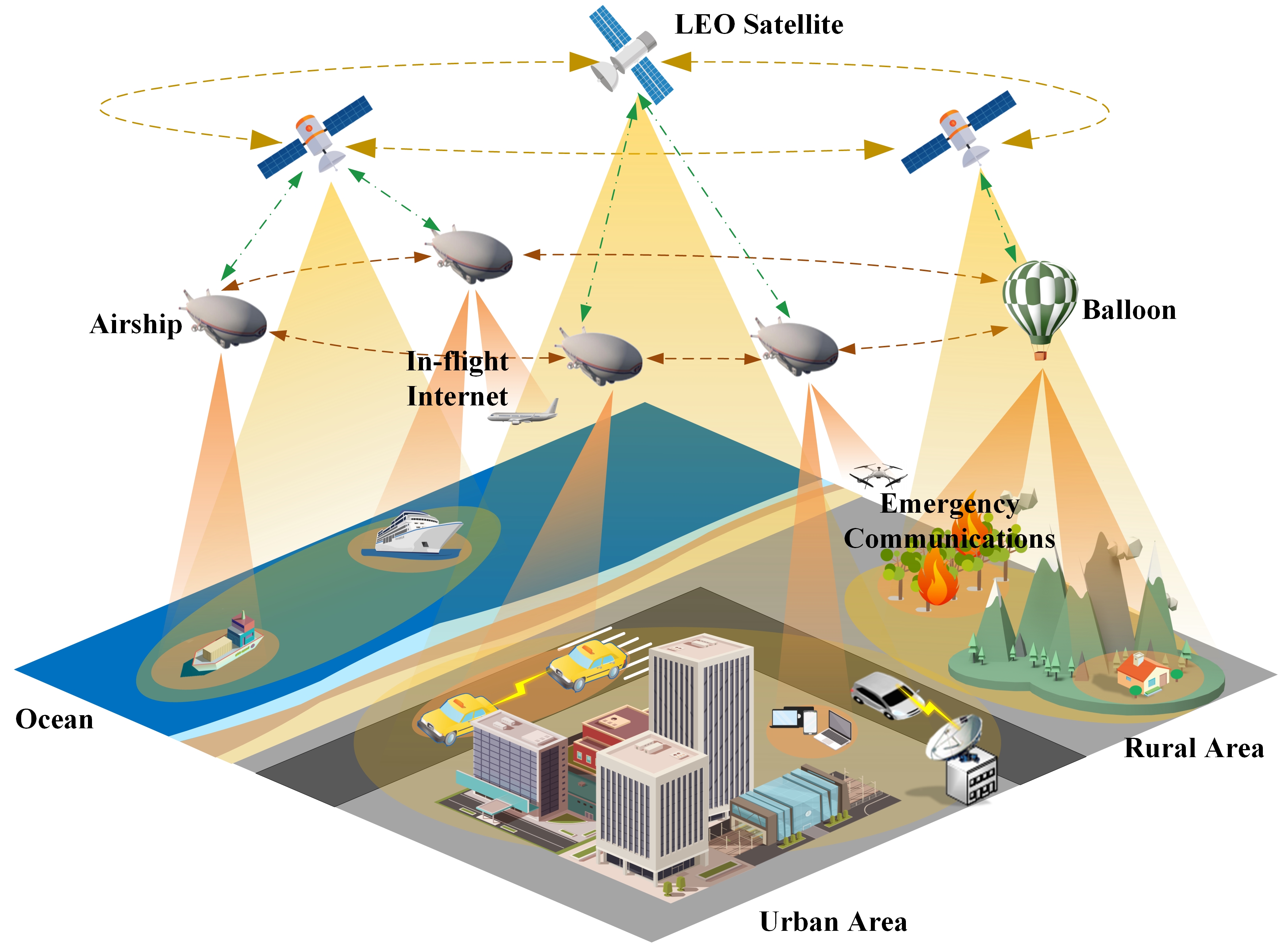}
\caption{Illustration of the scenario of NTN networks constructed by HAPS and LEO satellite constellations.}
\label{fig:NTN}
\end{figure*}

To provide broadband data access for transportation in distant regions, 
the NTN is envisioned as a promising complement to the existing ground network architecture, which is constructed by the heterogeneous nodes including the low earth orbit (LEO) satellites, high-altitude platforms (HAPS) and the unmanned aerial vehicles (UAVs), as shown in Fig. \ref{fig:NTN}. It is plausible that 6G will unify the terrestrial network (TN) and NTN \cite{Di_Cjiang_IoTJ}. Compared with the UAVs, HAPS can support broader coverage of $10^6$ $\text{km}^2$ with longer endurance, whilst incorporating the advantages of low deployment expenses and desirable throughput at the same time \cite{Arum_survey_20,Mao_WCM_22}. On the other hand, the existing LEO satellite constellations  are capable of providing nearly full earth coverage. Below we  discuss the roles of these typical NTN platforms in the future 6G network implementations and how these platforms can serve transportation in remote areas.

\subsubsection{HAPS as IMT Base Stations} 
HAPS is one of the effective methods to expand the wireless coverage over the ocean and for remote areas \cite{9198057}. According to the ITU Radio communication Sector (ITU-R) Radio Regulation (RR), HAPS is defined as the aircrafts (airship or balloons) operating at the altitude of 20-50 km, say, the lower near-space environment. Unlike the other aerial/spaceborne platforms with strong mobility, e.g., UAVs and satellites, the HAPS can keep quasi-stationary above the stratosphere for weeks and months. These make HAPS naturally qualified to provide stable LoS data access to terrestrial/aerial user terminals as IMT base stations, and such HAPS as IMT base stations (HIBS) are considered a good complement for 6G to provide coverage to remote areas and transportation such as ships on oceans. 

The unique near-space propagation environment  induces technical obstacles to the development of HIBS. Specifically, the temperature, pressure and constituent of the atmosphere present pronounced inhomogeneity across the lower near space, which causes sophisticated refraction/scattering/diffraction/absorption effects, adding difficulties in both channel modeling and beam alignment. 

{\bf Channel Modeling for HAPS-to-Ground Propagation}: Precise knowledge of the HAPS-to-ground communication environment can be mandatory for practical implementation of HIBS, which has motivated preliminary channel modelling works. In particular, two previous work  employs a non-stationary geometry-based stochastic model to characterize the HAPS-to-ground data links, where the dynamic property of the multi-path components is taken into consideration in \cite{Lian_tvt_19,Michailidis_tvt_20}. Besides,  a statistical channel model for the Ka-band HIBS systems by involving the weather effects and impairment of the ground environment is proposed \cite{Zhao_vtc_20}. These stochastic strategies no longer require the knowledge of the propagation environment, which can effectively alleviate the computational overhead at the cost of increased modelling errors. However, the existing literature fails to fully consider the complex factors of the HIBS communication channel. On the other hand, the AI philosophy has been invoked for communications to solve intractable problems for classical mathematical transceiving algorithms, such as signal detection under extremely sophisticated channel. Inspired by this, with sufficient channel measurement data for training, an artificial neural network is expected to implicitly characterize the relationship between the channel fading coefficients and the factors of HAPS-to-ground propagation environment, such as the mobility of the terminals/scatterers and the space-time-frequency varying refraction/scattering/absorption effects.

{\bf Terahertz Communications for HIBS Applications}: To realize ubiquitous coverage towards the 6G network, ultra-broadband wireless backhauling can be especially crucial between HIBS constellations. Classical microwave frequencies can hardly support inter-HIBS backhauling over 100 Gbps due to scarcity of the available bandwidth. Besides, the bulky parabolic antennas are likely to be employed for long-range backhaul over 100 km, which is contradicted to the limited budget for HIBS-mounted communication payloads. On the contrary, the emerging THz spectrum ranging from 0.1-10 THz is capable of providing Tbps data transmission with ``inexhaustible’’ frequency resources \cite{Elayan_ojcoms_20}. The size of THz-band antenna can be extremely undersized due to its short wavelength, say, with the length of 1.5 mm at 100 GHz. This enables ultra-massive antenna array that forms high-gain directional beams with compact size. In addition, different from the terrestrial environment, the inter-HIBS transmission above stratosphere experiences marginal path loss due to the sparse atmosphere. Therefore, THz communications are considered to naturally suitable for inter-HIBS backhauling in the near-space environment \cite{Mao_COMMAG_22}. Despite its merits, the employment of ultra-massive THz-band antenna arrays on HIBS may induce additional burden of excessive energy consumption, which is mainly originated from the large of RF chains and a sophisticated feeding network for beamforming. To tackle this issue, the reconfigurable metasurface technology can be invoked to replace the conventional phased-array antenna at the transmitter. Unlike classical RIS deployed between the source and destination nodes, the metasurface array can realize digital modulation and beamforming through flexible control of the reflection coefficient of the elements, without the need of RF chain and feeding network. This emerging RF-chain-free transmitter reduces both the hardware overhead and energy consumption, which is especially desirable for HIBS-mounted applications. 

\subsubsection{LEO satellite-based communication}

LEO satellite-based communication is especially useful to provide wireless connectivity for remote and ocean areas \cite{jamalipour1997low}. It is reported that the current wireless coverage with ground-based networks is limited to 15 km away from the coastline \cite{Di_SAGIN_TWC}. Albeit ship and aircraft-based relays within a heterogeneous network architecture can extend the wireless connectivity over the ocean to 60 km \cite{Di_Infocom, Di_LAPC}, it is still not enough to provide a wide coverage over the ocean, needless to say the wide broadband wireless services. To this end, LEO satellite-based communication is one promising solution to offer ubiquitous connectivity for the ocean and remote areas. Besides the wider coverage area, LEO satellite-based communication can also provide lower latency wireless connections compared to other satellite-based communications such as geosynchronous orbit (GEO).

In LEO satellite-based communications, the Iridium constellation was first convinced in 1987 to provide voice and data information coverage for satellite phone users. In 1989, Japan tested the satellite-based maritime communications between a sailing vessel and the ground earth station with a video telephone communication at 64 kbps through the Japanese Engineering Test Satellite (ETS-V) using the L band \cite{Di_ETS-V}. As of 1993, the first generation of Iridium constellation was developed, and its commercial services was subsequently provided in 1998, while a full global coverage was completed in 2002. As of 2017, the next generation of Iridium constellation was launched, and its commercial service was available in 2018 with a 704 Kbps speeds across maritime, aviation, land mobile, government, and IoT applications. As of 2012, OneWeb was founded, aiming at launching about 2000 LEO communication satellites to provide the global internet services especially for the unconnected and underserved communities. As of 2016, China launched the Tiantong LEO communication satellites, which can provide wireless services with 384 Kbps data rates. And in 2019, Amazon introduced the Kuiper project with 3236 satellites planned to launch to provide global, fast and low latency internet services. Starlink, another  LEO communication company, provides wideband wireless services globally including the ocean and remote areas, with ideal quality of experience (QoE) performances, i.e., 50-200 Mbps expected downlink speeds, 10-20 Mbps expected uplink speeds, and 99 ms average latency. LEO satellite-based communication is generally hampered by the Doppler shifts due to the high relatively moving speed between satellites and satellites to the ground receivers. Apart from that, enhanced broadband inter-satellite and satellite to earth communication technologies are needed in 6G to provide  better global services.

{\bf Effective technologies to combat the Doppler shifts}: A wireless communication transceiver moving at high speed introduces a significant Doppler effect that can adversely impact system performance. To address this challenge, a satellite can employ a proactive approach in the uplink by pre-compensating the frequency offset using a value estimated in the downlink. This strategy ensures that the satellite receives the uplink signal without a frequency offset, thereby substantially mitigating the Doppler effect. Similarly, in the downlink, the satellite can employ a comparable frequency offset estimation method for Doppler compensation. The iterative application of estimation and compensation in both the uplink and downlink enhances reliability. This iterative process can also benefit from prior knowledge, such as satellite trajectory and real-time heading information, further optimizing the compensation mechanism. Looking ahead to 6G, there is an anticipation of allocating higher frequency bands to provide expanded bandwidth compared to 5G. This increased bandwidth is expected to enhance estimation accuracy and enable more effective Doppler compensation, reinforcing the robustness of the overall communication system. Apart from that, OTFS modulation \cite{Di_OTFS_TWC}, affine frequency division multiplexing (AFDM) \cite{Di_AFDM} and resource sparsity-based technologies that does not rely on the orthogonality between neighboring carrier frequencies, such as non-orthogonal multiple access (NOMA), are emerging technologies to combat the Doppler's side effect.

{\bf Combination of wireless and free space optical communications}:  As the number of connected devices via wireless networks continues to rise, the scarcity of frequencies poses a formidable challenge, if not the most critical one, for achieving comprehensive wireless coverage through satellite communications. Additionally, reports indicate that adopting higher frequencies, such as mmWave and THz, for direct connections from cellphones to satellites is exceedingly challenging. To address this, widely accepted frequency reuse technologies will be integrated into NTN. Simultaneously, the integration of free space optical (FSO) communications alongside RF communications becomes imperative for heterogeneous TN and NTN networks. For instance, in systems like Starlink, laser communications are employed for inter-satellite links, effectively reducing latency, and the use of narrower laser beams eliminates interference, ensuring higher transmission security. Notably, for data transmission from Starlink satellites to ground users, carrier frequencies ranging from 10.7-12.7GHz and 37.5-42.5GHz are adopted. This strategic combination of technologies aims to overcome frequency shortages, optimize network performance, and enhance the overall efficiency of satellite communication systems.

\subsubsection{Standardization efforts for NTN}

NTN, as an umbrella term of any type of networks that involved of non-terrestrial objects, has been an active research filed in the 3rd generation partnership project (3GPP). In literature, wireless service continuity, ubiquity, scalability are the main concerns in 3GPP. The initial studies about NR-NTN in 3GPP was from Rel-15 with a focus about the deployment and channel modeling issues. Solutions for NR technologies to support was discussed in Rel-16, and a new work item about NTNs was started in Rel-17, to discuss the enhancements for NTN, which not only includes the LEO and GEO-based satellite communications, but also targeting other types of NTN networks such as HAPS \cite{Di_CSM}. It is widely agreed that NTN is capable of transparent forwarding information in Rel-18, and will be able to establish a satellite-based core network with BS on the fly in Rel-19.  The NTN is included as an indispensable part of 5G new radio (NR) in TS 22.261 \cite{Di_22.261}. In radio access network (RAN) work group 1 (RAN1), the use cases and service requirements are discussed in TR 22.822 \cite{Di_22.822}. Doppler and propagation delay are deemed as challenging physical layer issues for NTN, and channel models are provided by 3GPP \cite{Di_38.811}. In addition to RAN1, user plane and data plane issues in RAN2, network architecture and singnaling issues on integrating the TN and NTN in RAN3, are also widely discussed in 3GPP. The standardization activities of NTN are approved by 3GPP in Release 17 (Rel-17), and a NR-NTN and IoT-NTN enhancements are included in Rel-18. Although NTN has been widely acknowledged in 3GPP, yet the current discussed are mostly limited to the satellite-based communications, especially the LEO communications \cite{Di_CSM}. It is thus expected that other NTN types, such as HAPS and HIBS, will be extensively included in Rel-19 and the following releases.

In addition to a series of 3GPP discussions, the ITU has also conducted extensive discussions on the NTN technology. For instance, ITU-R study group 4 (SG4) is dedicated to satellite communications and the deliverable ITU-R M.2460-0 outlines how satellite communications are capable of supporting various wireless service scenarios, providing envisaged use cases and key technical elements while integrating the satellite solutions into next generation access networks \cite{Di_2460}. 
Additionally, other standard development organizations (SDOs) are also actively engaged in coordinating and harmonizing TN and NTN networks. For instance, the European telecommunications standards institute (ETSI) published TR 103 611 on satellite earth stations and systems \cite{etsi_103_611}.

\section{Intelligent transportation {empowered by 6G}}  \label{Sec_3}
On top of providing ubiquitous connectivity to passengers and devices on all types of transportation, 6G is also able to digitize them and empower intelligent and efficient transportation. Supported by the six usage scenarios of 6G, many transportation systems can benefit from not only the communication capability 6G offers, but also the beyond communication scenarios as well.

\subsection{High accuracy localization}
\subsubsection{Localization approaches}
Accurately estimating real-time positions such as vehicle location, UAV location, ship location, etc, is essentially  important for intelligent transportation. For example, traffic monitoring systems can intelligently control the stoplights as well as give alternate routes for motorists to avoid the traffic jam  when they know the accurate location  of the vehicles. 
As 6G is planned to be rolled-out in most urban regions around the world, 6G networks can satisfy localization needs from the traffic.
To accurately estimate the locations of vehicles, several classical wireless localization techniques can be used for communication and broadcasting networks, namely, the received-signal-strength (RSS) based method \cite{patwari2005locating}, time-of-arrival (ToA) based approach \cite{8421080},  angle-of-arrival (AoA) based approach \cite{8450280} and time-difference-of arrival (TDoA)  based method \cite{liu2019TDoA}. 
For the RSS-based localization technique, the receiver determines the range of the target based on the received  power strength of the echo signals, which typically rely on the free space physical model given as 
\begin{align}
P\left( d \right) = {P_0} - 10\alpha \log \frac{d}{{{d_0}}},
\end{align}
where $\alpha$ is the path-loss exponent and ${P_0}$ represents the received power at the  reference distance ${d_0}$. 
The ToA-based localization technique
relies on the measurement of the signal travel time, and the range of the target  can be easily computed as $d = c\tau /2$, where $c$ represents the transmission speed of the radio-frequency signals and $\tau$ denotes the time of transmission.  
TDoA based methods utilize the difference of the arrival instance of positioning signals from different Txs, thus, can be realized without strict time synchronization between the Rx and the Tx.
The above three techniques provide promising solutions for the range estimation, while the AoA-based localization technique, provides localization information  complementary to the former  techniques by providing the direction.  By equipping with a large number of antennas at the transmitter, the AoA is estimated from the differences in arrival times for a transmitted signal at each antenna. 
To acquire a higher accuracy, a combination of the above techniques is beneficial. For example, some literature \cite{taponecco2011joint,kim2017aoa,sun20163d} study target localization based on the joint  estimation of ToA and AoA under different system setups and demonstrate the accuracy estimation results. However, the above techniques are only applicable to the cases where the target is stationary or moving slowly. For the  case where the vehicles have higher speed, the vehicle's speed  must be considered to compensate for the system performance caused by the Doppler effect.  The key idea of estimating vehicles' speed  is comparing the difference between the transmit waveform frequency and the received echo waveform frequency \cite{kellner2013instantaneous}. 

\subsubsection{Improvement of localization accuracy}
The localization accuracy is affected by many conditions such as the bandwidth of the reference signal, channel condition, distance, and the transmit power.

6G is envisaged to be allocated with larger frequency bands than any previous generation, and since the estimation accuracy essentially relies on the bandwidth of the reference signal, it is estimated that 6G can provide a positioning accuracy of 1 to 10 cm, depending on the detailed configuration.

On the technical side,  several 6G technologies  can also improve the accuracy  such as ultra massive MIMO and cooperative sensing. When the transmitters or the receivers are equipped with a large number of antennas, the beamforming gains and the  diversity gain can be increased significantly. In addition, by carefully dispatching  multiple receiving  nodes and transmitting nodes in the area of interest,   more multiple signal  paths can be illuminated.

\begin{figure}[ht]
\centering
\includegraphics[width=0.95\linewidth]{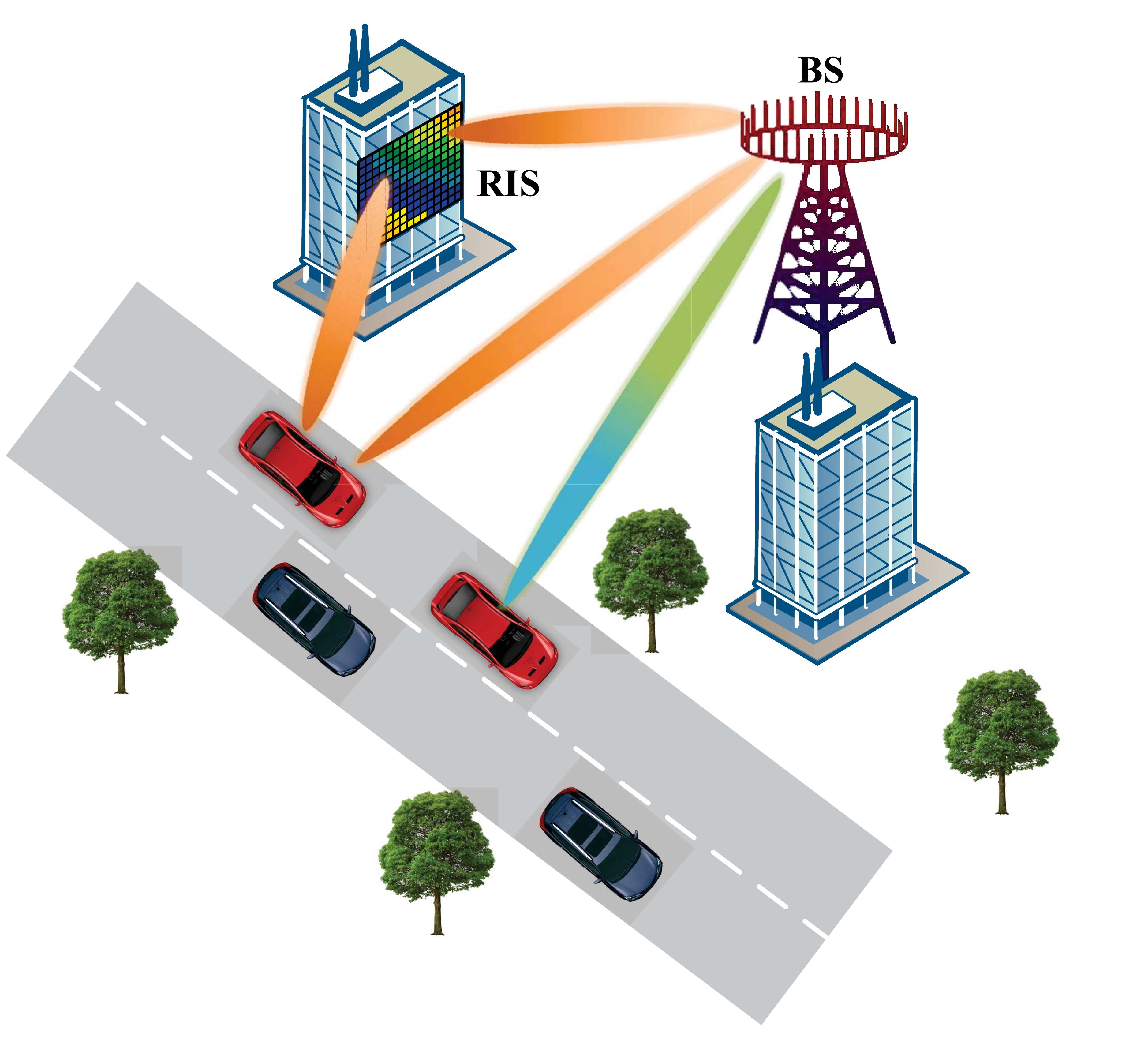}
\caption{Illustration of RIS-assisted localization enabled by 6G networks.}
\label{sensing_fig1}
\end{figure}

One major challenge of high accuracy localization for transportation system is the lack of LoS channel. This is due to the fact that transportation systems always come with mobility issues, and the available frequency bands for 6G will be higher than previous generations.
In transportation networks,  the wireless channel is usually blocked by complex environment objects such as buildings and trees in the urban area. To tackle this issue, one candidate 6G technology, RIS, is seen as a game changer since it can create a virtual LoS bypassing the transmitter and the target \cite{shao2022target}.  
The following advantages of RIS for sensing and localization can be achieved. 
Firstly, when the direct link is absent, the  RIS can enhance the energy of the echo signals and naturally improve the estimation accuracy  of the vehicles since the RIS can reconstruct the desired signal propagation environment by dynamically adjusting the amplitude and/or  phase of the incident signal. Secondly,
additional effective LoS reflection paths can be added to the detection target, thus providing more information about the target and helping to improve the accuracy and range of the perception. What's more, by configuring a small number of low-cost sensors on the RIS side, the RIS itself can also process data  based on the echoes it receives, thus increasing new degrees of freedom for estimation and detection, and the localization accuracy can potentially be enhanced by combining with the echo fusion processing of the traditional base station.
In Fig.~\ref{sensing_fig1}, we envision RIS-assisted localization enabled by 6G networks. It  can be seen that the vehicles in the shadow region that are blocked by the building can still be illuminated by the RIS. To maximize the sensing performance of the networks, the RIS deployment, RIS phase shift, and resource (such as time and frequency band) should be jointly optimized. Several optimization techniques can be applied such as standard convex optimization and machine learning techniques \cite{NAWAZ2023101976}. In particular, compared to the convex techniques that rely on the exact physical model as well as exact channel state information, machine learning-based techniques are applicable to  a complex environment that lacks environmental information, and  combining  partial physical environment information and  the experience/prior channel models, a satisfactory system performance improvement can be achieved \cite{lary2016machine}. 

\subsection{Integrated sensing and communication}
It is common knowledge that communication signals can be used to achieve sensing purposes, even by uncoordinated receivers \cite{9417324}.
In the current wireless network, communication and sensing usually rely on independent hardware systems that could form a coexistence of radar and communication (CRAC) system, to work independently in different ways \cite{zhang2021overview}. 
However, these systems face several challenges:  \emph{1) The signaling overhead of the channel estimation and information exchange is high:} In the CRAC system, it is  necessary to estimate not only the signal information between the communication base stations and users but also the channel between the  radar Tx and  users  as well as communication base stations.  In addition, some information exchanges are needed between the radar system and the communication system, such as the exchange of channel information, communication data, communication modulation mode, radar waveform, etc., which greatly increases the system signaling overhead. 
\emph{2) The interference is severe:} On one hand, the communication user will receive the sensing signal from the radar. On the other hand, the echo signal received by the radar not only contains the information of the target but also contains the signal transmitted by the communication base station.  Therefore, co-frequency interference will restrict the performance limit of CRAC systems.
\emph{3) The cost of hardware is high:} In order to efficiently utilize the same spectrum resources to improve overall system performance, additional unit controllers need to be designed to collaborate with radar systems and communication systems.  At the same time, due to the strict requirements of signal synchronization for communication and sensing functions, additional precision equipment such as the phase-locked loop is needed, which dramatically increases the hardware cost. 

In order to solve the above challenges, the concept of ISAC is proposed \cite{liu2022integrated} and has quickly become popular and been recognized as a key 6G usage scenario. By integrating the sensing and communication functionalities into a common hardware platform, resource utilization is significantly improved so that the wireless network can not only carry out high-quality communication but also provide high precision for sensing \cite{10233791}.  To be specific, the following advantages can be achieved:  \emph{ 1) Information exchange signaling overhead is low:} Since the communication function and radar sensing function are integrated on one platform, the number of interference links of the overall system is greatly reduced, so the interference is sharply reduced and the signaling overhead for information exchange  only needs to be coordinated at the integrated receiver, there is no need to design a special coordination control unit. \emph{2) Integrated gain is high:} As the communication and radar resources are integrated into a common hardware platform, the optimization of resources such as time and frequency could more effectively achieve better system performance trade-offs. \emph{ 3) Cooperation gain is high:} Sensing  and communication can help each other.  On the one hand, the received echo signals contain part of the target information, such as the target's position and direction, which  greatly reduces the beam training overhead for communication.  On the other hand, the node with limited computing resources can send the sensing  information to the server with powerful computing functions such as edge base station through wireless communication transmission, and use the server to quickly parse the target state information. 

ISAC can already be realized by 5G networks as recent field trials demonstrated \cite{liu2023ISAC,10182512}, while its true power will be fully unleashed by 6G.

\subsubsection{Resource allocation management in ISAC systems}
In contrast to the CRAC systems, the ISAC systems can simultaneously send communication signals and sensing  signals at the dual-functional base stations. Therefore, the  joint design of  sensing  waveform and communication waveform  is needed  by taking account into both data transmission rate and sensing performance. In general, sensing  waveform and communication waveform can be classified according to different service application scenarios to reduce hardware costs. For example, for the sensing-centered integrated system applicable to the internet of thing (IoT) devices requiring low transmission rates, the communication information, such as radar signal modulation, radar bourbon modulation, and index (radar antenna) modulation, can be directly embedded in the sensing signal to achieve the purpose of communication, while significantly reducing the number of required radio frequency  links. For the communication-centered integrated system, target sensing  can be carried out directly with communication signals without adding additional sensing signals (however, because communication signals are random, the sensing ability is weak). For scenarios requiring a high transmission rate and high sensing  accuracy, the communication waveform and sensing  waveform can be jointly designed and optimized by using efficient optimization algorithms, such as the penalty function method, continuous convex approximation method, semi-positive definite relaxation method, etc., so as to meet the communication transmission rate and sensing performance at the same time. For example, some researchers studied the joint waveform design to maximize the sensing performance while satisfying the minimum achievable rates for communication users  \cite{liu2020joint}, and a low complex algorithm, namely, zero-forcing based algorithm, was  proposed.  There are also studies to further answer whether radar waveforms are needed in the joint waveform design approach under different receiver types at the users \cite{hua2023optimal}. A unified channel model for both the communication and sensing purposes \cite{Lou_vtc_23} would also be beneficial to facilitate the resource allocation in ISAC systems.

\begin{figure}[ht]
\centering
\includegraphics[width=0.9\linewidth]{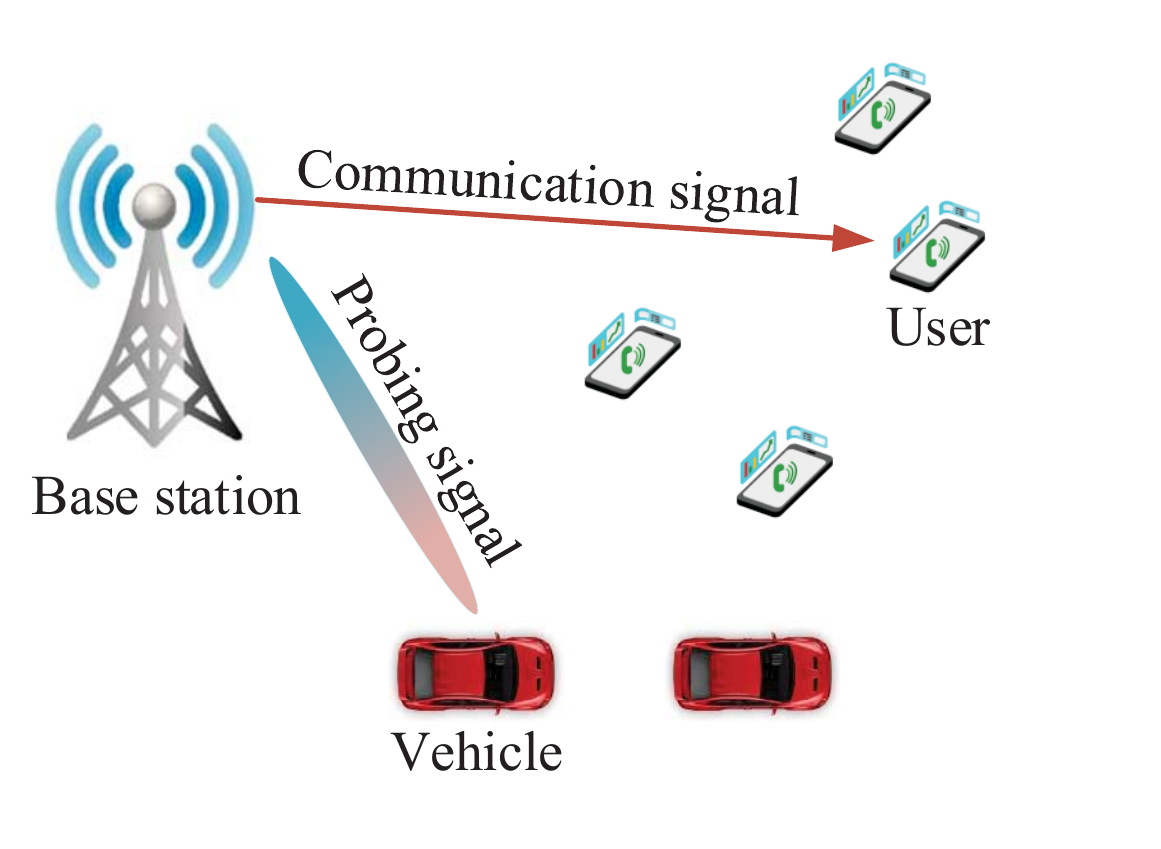}
\caption{Illustration of an ISAC system for transportation.}
\label{ISAC_fig}
\end{figure}

Although the ISAC systems provide a promising approach to simultaneously realize sensing and communication in an efficient way, they suffer an asymmetric communication and sensing coverage issue. To be specific,  as the sensing signal needs to go through the secondary reflection on the target object, the attenuation is larger than the communication signal, so the sensing  distance of the base station might be less than the communication distance.  Besides, the sensing system is more dependent on the LoS link between the base station and the target, and the obstacles such as buildings  severely limit the sensing range, while the non-line-of-sight (NLoS) signal  reflection can also improve the communication performance.  The above challenges limit the development and application of ISAC, which needs to be further studied and solved.  Fortunately, the RIS is beneficial for both sensing and communication.  In fact, the RIS is not only can create virtual LoS links for sense but also can  create more NLoS channel conditions for multi-user communication to improve the rank of the multi-user channel matrix and thus multiplexing gain is achieved. With the introduced RIS, the uncontrollable  electromagnetic wave propagation in the ISAC systems can be adjusted as possible. An illustration of RIS-empowered ISAC systems is presented in Fig.~\ref{ISAC_fig}. The sensing waveforms, communication waveforms, and RIS phase shifts can be jointly designed to cater to different quality service requirements.  For example, the authors in \cite{9913311} studied joint optimization of waveforms, communication waveforms, and RIS phase shifts to minimize the total transmit power subject to the minimum signal-to-interference-plus-noise ratio (SINR) required by communication users and  the minimum SINR required by the radar, and proposed a penalty-based algorithm and semi-definite relax-based algorithm to cater to different constraints. The research of the ISAC is still ongoing, and there are open issues that need to be addressed such as the optimal deployment of the base station, the coordination of base stations and novel design of system architecture and protocols.

\subsubsection{Sensing-assisted communication}
The  ISAC  can realize unique sensing-enabling communication functions, which can be applied to more diversified and unique scenarios. For example, in the auto-driving scenario, due to the high-speed movement of vehicles, frequent channel requests and estimates, not only dramatically increase the signaling overhead, but also  the current estimated channel information is not applicable to the next moment of information transmission due to  the fast fading characteristics of the channel. To solve this problem, different from the traditional pilot estimation method, the beam of the dual-functional base station can be used for joint training. Specifically, the dual-functional base station first sends the detection signal and then uses the received echo signal (at the dual-functional base station) to conduct joint training/matching and optimization of the  codebook at the base station, and then uses the trained codebook as a communication beam for information transmission. In addition, in order to further reduce the beam search space, based on the currently obtained beam, the time/space correlation is used to predict the beam at the next moment (such as using the Kalman filter for prediction estimation), to achieve real-time ground beam tracking. The first work \cite{liu2020radar}  
investigated a radar-assisted predictive beamforming design for
vehicle-to-infrastructure communication in ISAC systems, and presented a novel extended Kalman filtering  framework to track and predict kinematic parameters of each vehicle. Then, an extended vehicle target for sensing-assisted communication was studied in \cite{9947033} and a novel sensing-assisted beam tracking protocol was proposed.

\subsection{Autonomous driving}
Talking about future transportation, autonomous driving might be the first thing pops up in many minds. Autonomous driving allows vehicles to navigate and drive, even fully without human intervention. There are different levels of automation to achieve, and almost all scenarios require communication features and capacities. Supported by communication features as well as beyond communication features such as localization and sensing, autonomous driving is gradually becoming a reality. To fully achieve autonomous driving, there are some new communication paradigms 6G need to support.

\subsubsection{V2X communication}
The vehicle-to-vehicle (V2V), or more generally speaking, vehicle-to-everything (V2X) communication, usually require ultra reliability and extremely low latency, which can be served by the usage scenario of extreme communication in 6G. As shown in Fig. \ref{fig:6G}, extreme communication emerges from the URLLC in 5G, taking one step further by meeting more stringent requirements as presented in Table \ref{Tab:6G}. Moving at a speed of around 120 km per hour on the highways, vehicles travel a distance of approximately 33.3 meters per second and a reaction within 50 ms can satisfy most cases. For other scenarios such as truck platooning or autonomous driving in urban areas, reaction time of 50 ms would also suffice  \cite{darbha2018benefits,zardari2022adaptive}. This latency is calculated from the moment the collected data from sensors or cameras is transmitted to the moment the signal containing proper decisions is received, and thus, depend on the physical distance between the vehicle and the service center. It also depends on the processing capacity of baseband units in both transmitters and receivers as well as the calculation capacity. If the decision is made by collecting data from more than one vehicles and roadside infrastructure, the situation is more complicated and the reaction time becomes a more stringent requirement.
With all the above factors being considered, 6G is also needed to provide an extremely low air interface latency of around 0.1 to 1 ms, to allow sufficient information exchange among all the necessary nodes and to navigate the vehicles safely and more efficiently.

On top of the latency requirement, high reliability is also needed for V2X communication. This issue has been discussed in the era of 5G, where dedicated and enhanced HARQ processes were introduced \cite{haroundabadi2020v2x}. As 6G targets to provide stronger support for vertical industries such as transportation, it is expected that the reliability will be further improved, potentially reaching a new level of $1-10^{-6}$.

One difference between 6G and previous generations is that 6G is expected to be designed in a way that balance downlink and uplink transmission, as more and more use cases demand heavy uplink bandwidth,  V2X being one of them. The typical data generated by sensors on a vehicle per second is summarized in Table \ref{Tab:V2X}, which reflects that the user experienced uplink data rate may need to reach  750 Mbps for typical cases, if all the data is to be transmitted to service centers on the cloud. This exceeds the typical capacities of 5G, and will need to be satisfied by 6G.

\begin{table}[t]
     \caption{Typical data volume generated by sensors on vehicles}
     \label{Tab:V2X}
    \centering
    \begin{tabular}{|l|l|}
    \hline
        \bf{Sensor type} & \bf{Data volume per second}  \\ \hline
        Cameras & 160 - 192 Mbps  \\ \hline
        Lidar & 80 - 560 Mbps  \\ \hline
        Sonar & less than 1 Mbps  \\ \hline
        Radar & less than 1 Mbps  \\ \hline
        GPS & less than 1 Mbps  \\ \hline
    \end{tabular}
\end{table}

For most autonomous driving scenarios, mobility usually range from several km/h to no more than 200 km/h, which can be readily satisfied by both 5G and 6G and will not pose major challenges to cellular network enabled autonomous driving.

\subsubsection{Mobile edge computing}
There are several ways to navigate the vehicles in real time, for instance letting the on-board computers to make decisions locally, or, to upload data to a server and receive the instructions. As the future transportation system may comprise of many nodes, such as other vehicles, traffic lights and roadside sensors, it is reasonable for a server to collect all the information and make decisions comprehensively. However, making decisions at the service center might increase the end-to-end (e2e) latency, which depends on the physical distance between the nodes and the service center. To alleviate this problem, mobile edge computing (MEC) is introduced into V2X communication to allow calculating and making decisions near the vehicles, decreasing the e2e latency significantly. The 6G base stations and roadside units (RSUs) can potentially provide computational resources for the tasks, and at busy times, parked cars nearby can also take some of the computational tasks \cite{qi2021extensive}. Integrating computational capacities in base stations, 6G can better coordinate vehicles on the road and help to navigate them autonomously. The challenges of MEC for autonomous driving include smart scheduling of edge resources, dynamic resource allocation for MEC nodes and harmonizing MEC with other candidate 6G technologies \cite{9982493}.

\subsubsection{Network slicing}
To empower autonomous driving, the  aforementioned features will all need to be supported, bringing huge challenges to the network since these features have different demands for latency, bandwidth and reliability. Network slicing,  an innovative design principle that allows operators to create multiple logically independent networks (slices)  on shared physical infrastructure, can be of help \cite{khan2021end}. 

The network slices differ from each other not only by the time and frequency resources, but also by the QoS they support. Deployed on same physical infrastructure, one slice can serve extreme communication with very low latency while the other supports immersive communication with ultra-large bandwidth on the uplink.

To support totally different requirements of future transportation system using different slices, 6G needs to be designed from the beginning in a flexible way that is easy to be configured and managed. 
The key challenge is how to dynamically configure different RAN slices and how to manage them effectively in a timely manner, where AI empowers potential opportunities to solve the issue.

\subsection{Security for intelligent transportation systems}

\begin{table*}[t]
  \centering
   \caption{\label{ITU_T_17}Summary of ongoing work items on ITS under ITU-T SG 17}
	\begin{tabular}{|p{66pt}|p{160pt}|p{195pt}|}
		\hline
		\bf{Work item} & \bf{Title} & \bf{Summary} \\
		\hline
		X.1373rev & Secure software update capability for intelligent transportation system communication devices & Specifying a basic model of remote software update, and software update procedure for outside and inside of the vehicle, respectively. \\
		\hline
		X.evpnc-sec & Security guidelines for electric vehicle plug and charge (PnC) services using vehicle identity (VID) & Identifying security threats and requirements for electric vehicle plug and charge service.\\
		\hline
		X.evtol-sec & Security guidelines for electric vertical take-off and landing (eVTOL) vehicle in an urban air mobility environment & Stressing security threats, requirements, technical and environmental consideration of connected eVTOL systems used as an urban air mobility.\\
        \hline
        X.idse & Evaluation methodology for in-vehicle intrusion detection systems &  Providing a methodology for evaluating in-vehicle intrusion detection systems and identifying its functional, performance, security and deployment requirements.\\
	\hline
        X.itssec-5 & Security guidelines for vehicular edge computing & Analyzing threats and vulnerabilities of vehicular edge computing, and providing use cases and related security requirements for such systems.\\
        \hline
	    X.sup-cv2x-sec & Supplement to X.1813 - Security deployment scenarios for cellular {vehicle-to-everything} (C-V2X) services supporting URLLC  &  Identifying security threats and describing security deployment scenarios for C-V2X services supporting URLLC.\\
        \hline
       X.ota-sec & Implementation and evaluation of security functions to support over-the-air (OTA) update capability in connected vehicles &  Stressing security threats, requirements, technical and environmental considerations of connected vehicles which support OTA capabilities.\\
		\hline
	\end{tabular}
\end{table*}

ITSs have become more and more complicated, exposing itself to potential security threats. In this section, how 6G can provide security for ITSs is introduced.

In the academia, how novel technologies can be applied to secure future ITS is under study. 
The applications of block chain, machine learning, and quantum-related technologies in ITS are studied by many researchers, achieving ideal performance in improving the security of ITS \cite{chulerttiyawong2021blockchain,chulerttiyawong2023sybil,QKD_security,gary2023Quantum}.
Besides those academic works, several SDOs devote themselves to security issues of ITS and make recommendations providing guidelines for ITS security.
In the ITU Telecommunication Standardization Sector (ITU-T), study group (SG) 17 works on security-related standardization issues including the security of applications and services for ITS. 
There are several work items concluded their work and published recommendations on enhancing the security of ITS.
The first recommendations focus on identifying the security threats in ITS, where the security threats to connected vehicles and V2X communication are stressed in work items X.1371 and X.1372, respectively. 
Work items X.1373 to X.1377 provide implementation guidelines to give instructions on preventing security threats from different aspects, like establishing an intrusion detection system for in-vehicle networks and applying a misbehavior detection mechanism for connected vehicles.
Other work items propose security requirements for different functions in ITS, where requirements for roadside units are specified in X.1379 and those for categorized data are specified in X.1383.

Considering that the rapid development of networks brings ITS {potentially} new security threats and requirements, in the current study period, ongoing work items focus on revising existing recommendations according to arising requirements and developing new ones to improve security with emerging technologies.
In order to achieve thorough security of software update for ITS {devices}, work item X.1373rev makes modification on X.1373, which only stresses the update procedures outside vehicles by enlarging the scope of this recommendation to cover the in-vehicle update procedures.
Similarly, work item X.sup-cv2x-sec extends the content of X.1813 by adding deployment scenarios of vehicles connected to public 5G network, making the scenarios suitable for C-V2X service to meet the {5G URLLC requirement}.
Those revisions and supplements are believed to realize better coverage of security requirements in new usage scenarios as well as to provide more practical guidelines of implementation for relevant industries.
Besides, several work items aim to deal with security issues related to ITS supported by new technologies.
Vehicular edge computing is effective in decentralizing computing resources and achieving lower latency, yet brings security challenges like data leakage and data manipulation. To deal with the security threats and vulnerabilities, work item X.itssec-5 provides guidelines for {vehicular} edge computing, identifying the potential security problems and proposing security requirements and use cases for {vehicular} edge computing scenarios.
As the software-defined vehicle becomes a new trend {for the future}, software updating over the air brings vulnerabilities of cyberattacks due to a large amount of external interfaces. Work item X.ota-sec analyzes the potential vulnerabilities, security requirements and {implementations} of security functions, providing methodologies for automotive industries.
Details of the ongoing work items are summarized in Table \ref{ITU_T_17}.

\section{Conclusions}  \label{Sec_conclusion}
6G will change many aspects of the human society, including transportation.  Six key usage scenarios have been envisaged for 6G, including potential game changers for transportation such as massive communication, HRLLC, AI and communication and ISAC. 
By effectively overcoming the implementation challenges, transportation can benefit from many of the 6G features and capabilities, providing better overall services to passengers all around the world. Specifically, 6G can connect the previously unconnected devices on transportation such as high-speed trains and airplanes, as well as enable more intelligent, robust and secure transportation.

\bibliographystyle{IEEEtran}
\bibliography{sample}

\begin{IEEEbiography}[{\includegraphics[width=1in,height=1.25in,clip,keepaspectratio]{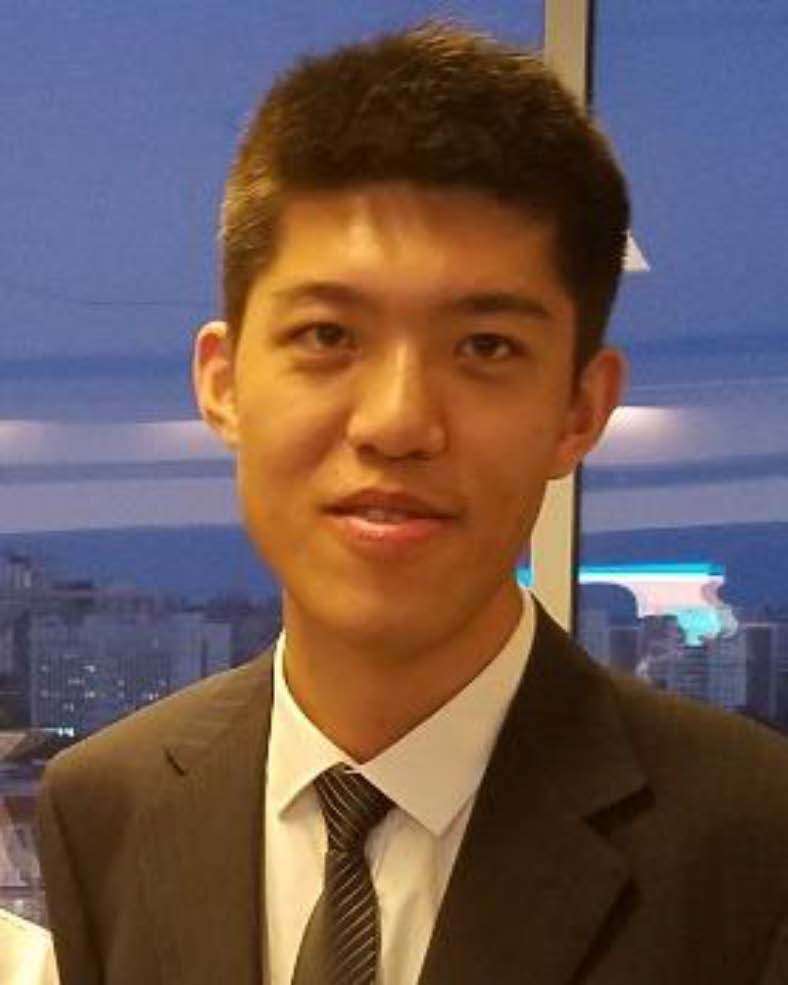}}]{Ruiqi (Richie) Liu} (S'14-M'20) received the B.S. and M.S. degree (with honors) in electronic engineering from the Department of Electronic Engineering, Tsinghua University in 2016 and 2019 respectively. He is now a master researcher in the wireless and computing research institute of ZTE Corporation, responsible for long-term research as well as standardization. His main research interests include reconfigurable intelligent surfaces, integrated sensing and communication and wireless positioning. He is the author or co-author of several books and book chapters. He has participated in national key research projects as the researcher or research lead. During his 3-year service at 3GPP from 2019 to 2022, he has authored and submitted more than 500 technical documents with over 100 of them approved, and he served as the co-rapporteur of the work item (WI) on NR RRM enhancement and the feature lead of multiple features. He currently serves as the Vice Chair of ISG RIS in the ETSI. He actively participates in organizing committees, technical sessions, tutorials, workshops, symposia and industry panels in IEEE conferences as the chair, organizer, moderator, panelist or invited speaker. He served as the guest editor for Digital Signal Processing and the lead guest editor for the special issue on 6G in IEEE OJCOMS. He serves as the Deputy Editor-in-Chief of IET Quantum Communication and the Editor of ITU Journal of Future and Evolving Technologies (ITU J-FET). He is the Standardization Officer for IEEE ComSoc ETI on reconfigurable intelligent surfaces (ETI-RIS) and the Standards Liaison Officer for IEEE ComSoc Signal Processing and Computing for Communications Technical Committee (SPCC-TC). His recent awards include the 2022 SPCC-TC Outstanding Service Award and the Beijing Science and Technology Invention Award (Second Prize, 2022).
\end{IEEEbiography}

\begin{IEEEbiography}[{\includegraphics[width=1in,height=1.25in,clip,keepaspectratio]{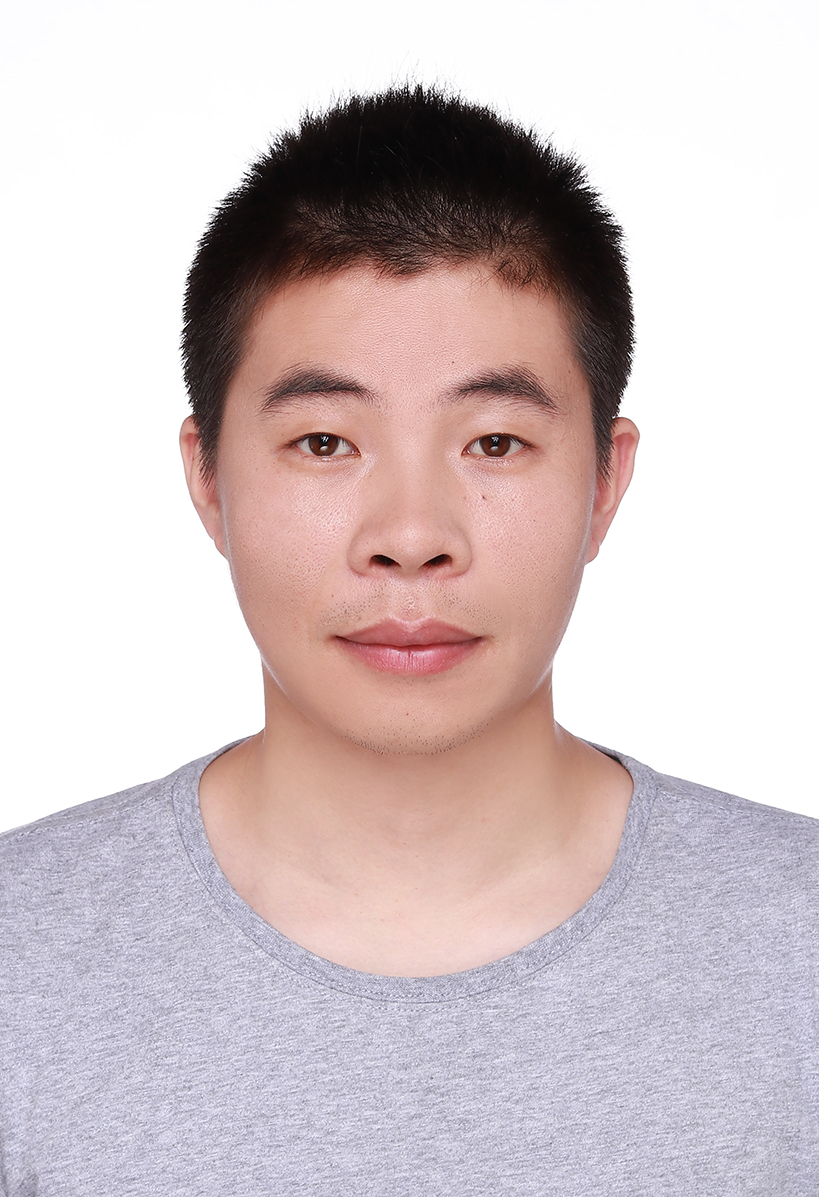}}]{Meng Hua}
	received the  M.S. degree in electrical and information engineering from Nanjing University of Science and Technology, Nanjing, China, in 2016, and Ph.D. degree in School of Information Science and Engineering, Southeast University, Nanjing, China, in 2021. He was the recipient of the  Outstanding Ph.D. Thesis Award of Chinese Institute of Electronics in 2021. From 2021/01-2023/06, he was a Postdoc at the University of Macau.  He is now a Postdoc at the City University of Hong Kong. He is an Associate Editor of the IEEE Open Journal of the Communications Society, Digital Signal Processing, and Physical Communication. His current research interests include localization,  integrated sensing and communication, and intelligent reflecting surface-assisted communication.
\end{IEEEbiography}

\begin{IEEEbiography}[{\includegraphics[width=1in,height=1.25in,clip,keepaspectratio]{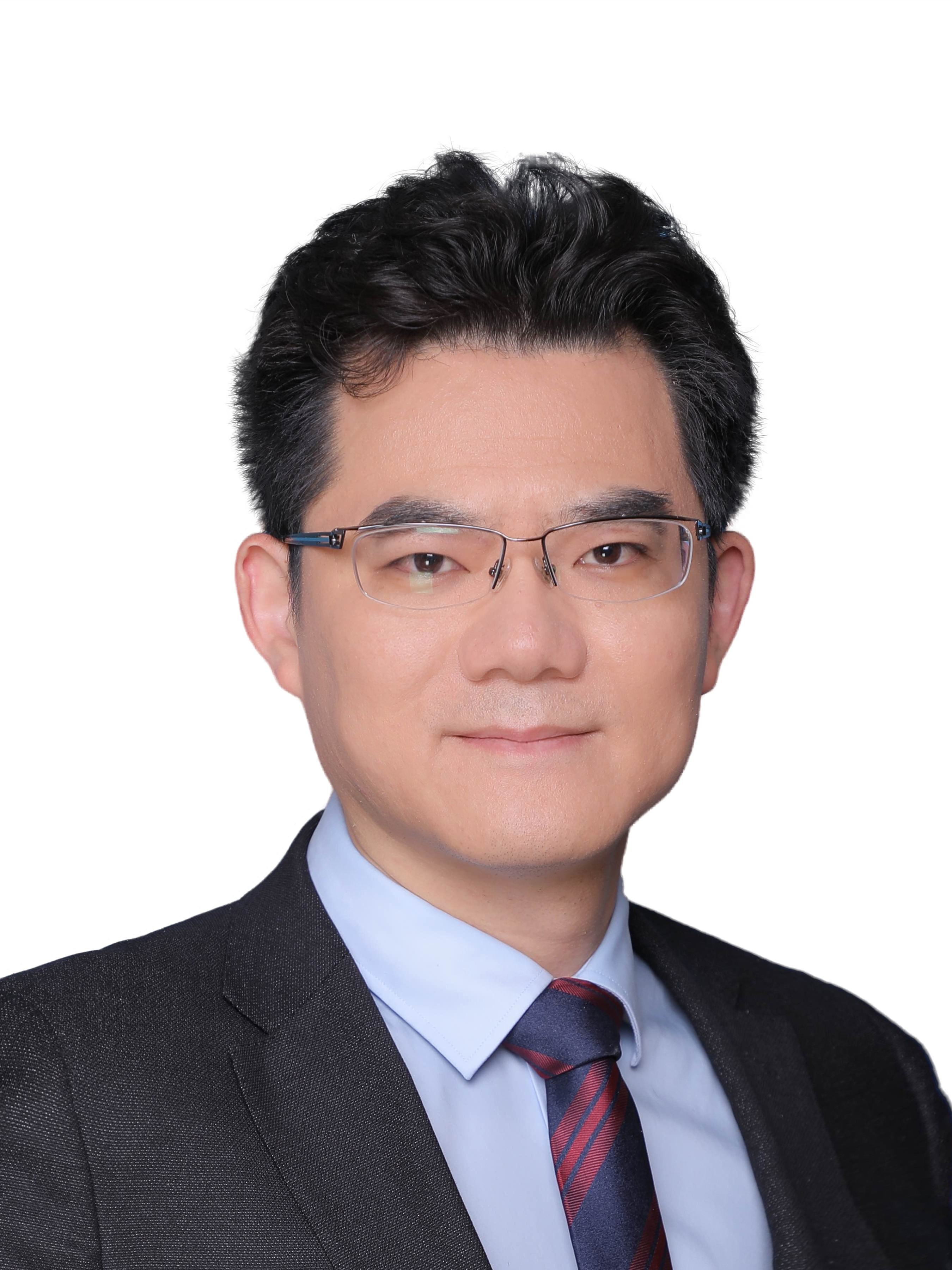}}]{Ke Guan} (S'10-M'13-SM’19) received B.E. degree and Ph.D. degree from Beijing Jiaotong University in 2006 and 2014, respectively. He is a Professor in State Key Laboratory of Advanced Rail Autonomous Operation and \& School of Electronic and Information Engineering, Beijing Jiaotong University. In 2016, he has been awarded a Humboldt Research Fellowship for Postdoctoral Researchers. He was a Visiting Scholar with Universidad Polit\'{e}cnica de Madrid, Spain in 2009 and 2013, respectively. From 2011 to 2013 and from 2016 to 2018, he was a Research Scholar with the Institut f\"ur Nachrichtentechnik (IfN) at Technische Universit\"at Braunschweig, Germany. From February 2023 to July 2023, he was a Guest Professor at Technische Universit\"at Wien, Austria. He has authored/coauthored two books and five book chapter, more than 200 journal and conference papers, and ten patents. His current research interests include measurement and modeling of wireless propagation channels, high-speed railway communications, ray-tracing and machine learning based digital twin of electromagnetic environments in various complex scenarios, such as vehicle-to-x communications, terahertz communication systems, integrated sensing and communications, and space-air-ground integrated networks.

Dr. Guan is the pole leader of EURNEX (European Railway Research Network of Excellence). He was the recipient of 2014 International Union of Radio Science (URSI) Young Scientist Award, 2023 Emerald Global Outstanding Award, and 2023 JIMSE Global Young Scientist Award in Advanced manufacturing. He is listed the World's Top 2\% Scientists (both for the single year 2022 and the whole career). His papers received 14 Best Paper Awards, including IEEE vehicular technology society Neal Shepherd memorial best propagation paper award in 2019 and 2022. He is an Editor of IEEE Vehicular Technology Magazine and IET Microwave, Antenna and Propagation, and a Guest Editor of the IEEE Transactions on Vehicular Technology and IEEE Communication Magazine. He serves as a Publicity Chair in PIMRC 2016, the Publicity Co-Chair in ITST 2018, the Track Co-Chair in EuCNC, the International Liaison of EUSIPCO 2019, the Session Convener of EuCAP 2015-2024, and a TPC Member for many IEEE conferences, such as Globecom, ICC and VTC. He has been a delegate in 3GPP and a member of the IC1004, CA15104, and CA20120 initiatives.

\end{IEEEbiography}

\begin{IEEEbiography}
[{\includegraphics[width=1in,height=1.25in,clip,keepaspectratio]{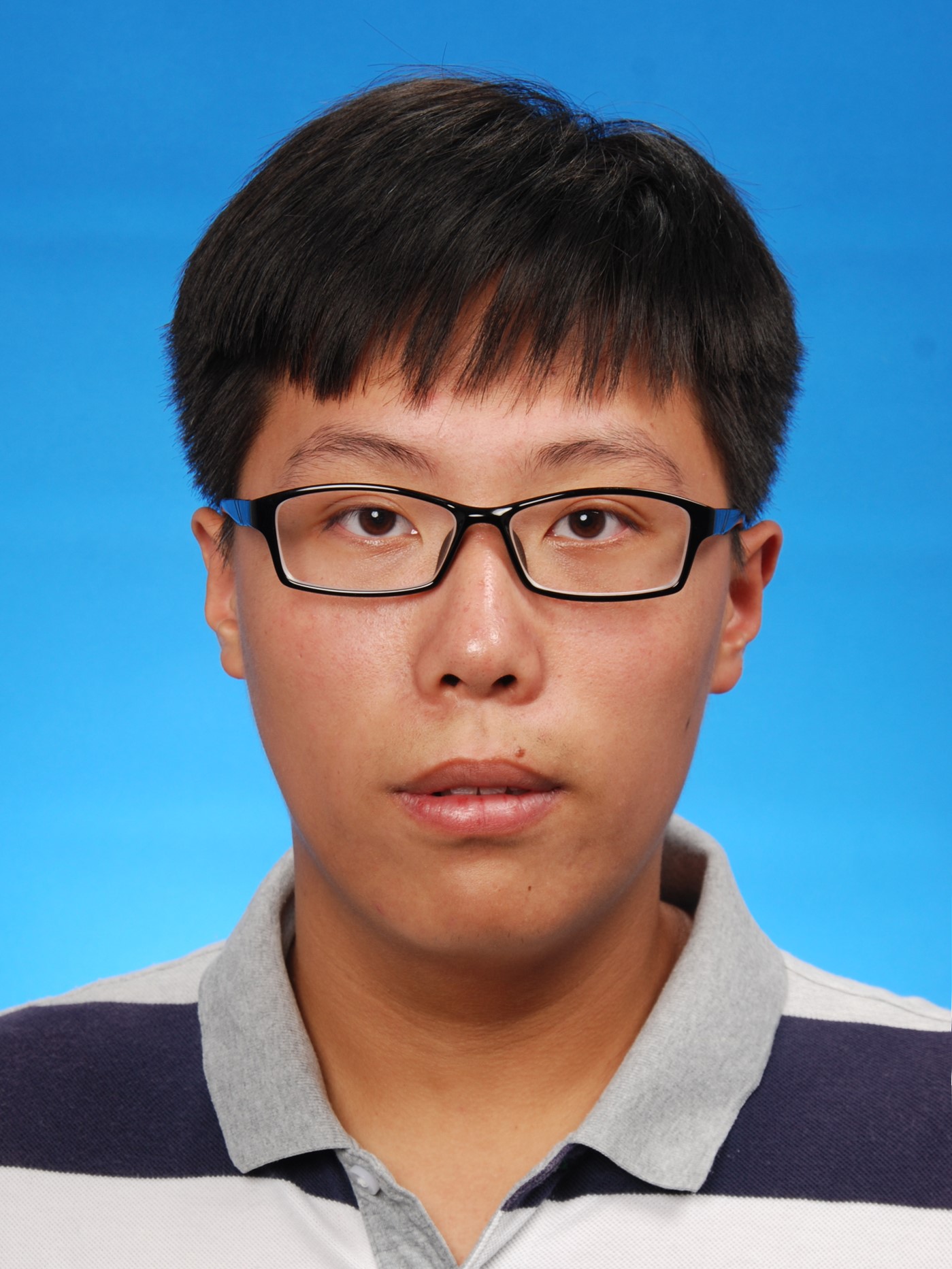}}] {Xiping Wang} (S'22)
was born in Qingdao, China. He received the B.E. degree from Beijing Jiaotong University, Beijing, China, in 2020. He received the M.Sc. degree of Electrical Engineering  from  National University of Singapore, Singapore. Currently he is working toward the Ph.D degree with State Key Laboratory of Advanced Rail Autonomous Operation, Beijing Jiaotong University, Beijing, China. He has a wide interest in research of wireless communication, including channel modeling, optimization, and modeling of wireless sensor network. Currently he is studying new models and algorithms for wireless channel modeling.
\end{IEEEbiography}

\begin{IEEEbiography}[{\includegraphics[width=1in,height=1.25in,clip,keepaspectratio]{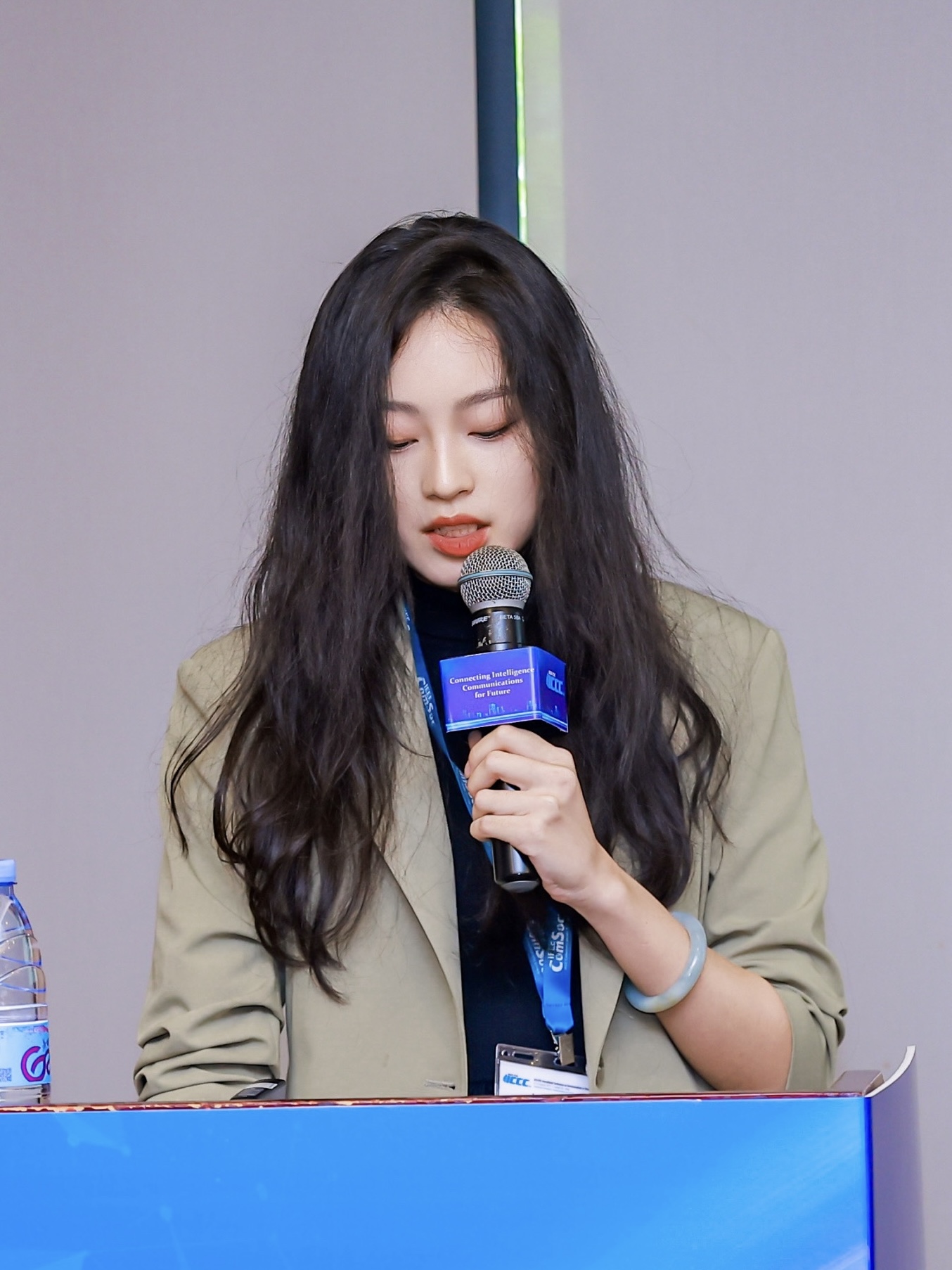}}]{Leyi Zhang} (S'22, M'23) received the B.S. and M.S. degree (with honors) in electronic engineering from the Department of Electronic Information Engineering, Beihang University in 2020 and 2023 respectively. She is working as a researcher with the technology planning department of ZTE Corporation, responsible for long-term research as well as standardization. Her research interests include network security and quantum communication. She has many peer-reviewed publications. She took part in many international academic conferences as the session chair or invited speaker. She participated in national key research projects as the research assistant or researcher. She actively participates in global standardization activities through platforms such as ITU-T SG 17, 3GPP SA3, ETSI and CCSA.
\end{IEEEbiography}

\begin{IEEEbiography}[{\includegraphics[width=1in,height=1.25in,clip,keepaspectratio]{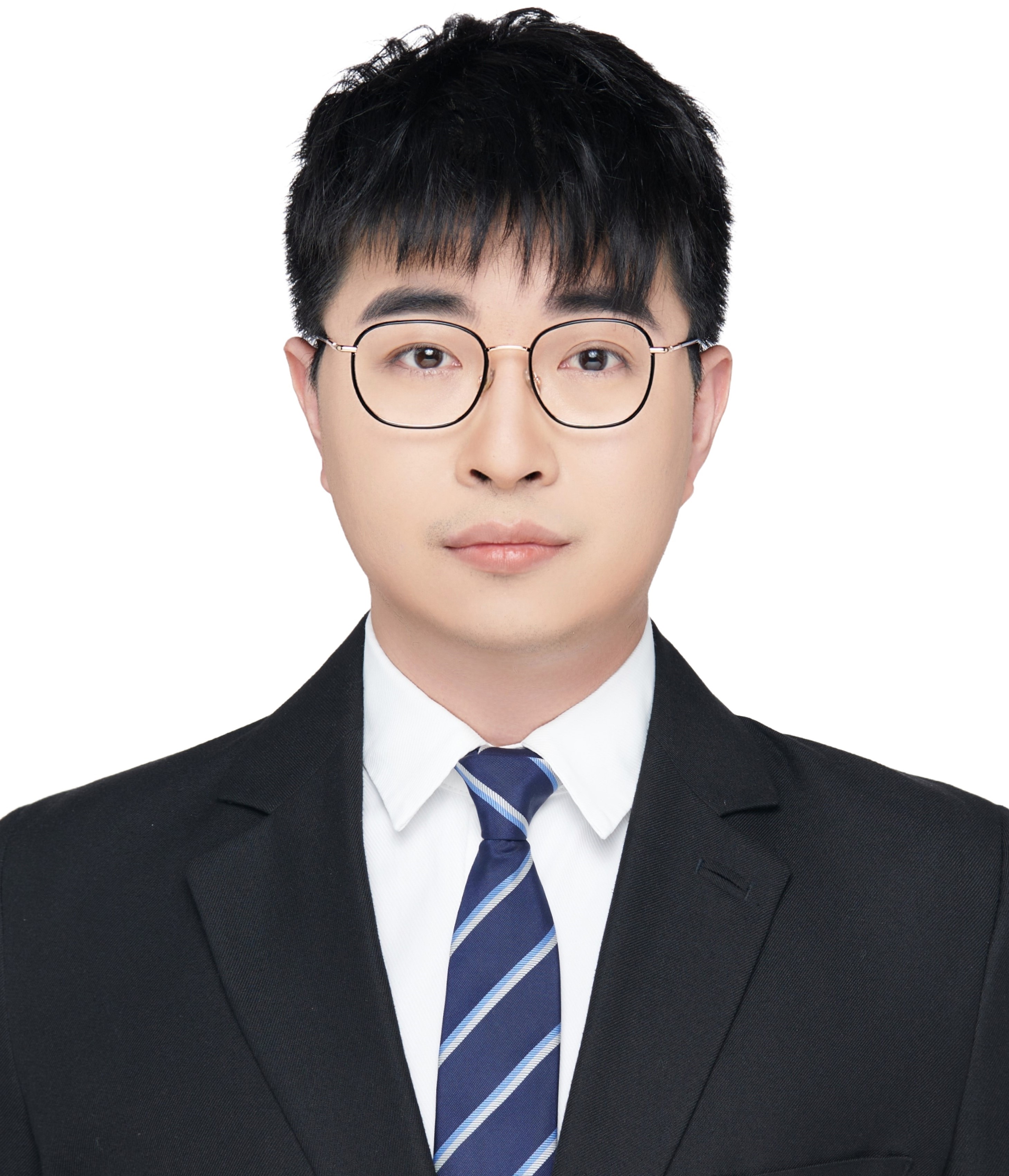}}]{Tianqi Mao} (M'22) received the B.S., M.S. (Hons.), and Ph. D (Hons.) degrees from Tsinghua University in 2015, 2018 and 2022, respectively. He is currently working with the Advanced Research Institute of Multidisciplinary Sciences, Beijing Institute of Technology, Beijing, China. He has authored over 30 journal and conference papers in IEEE COMST, IEEE JSAC, IEEE WCM, IEEE COMMAG, IEEE TCOM, IEEE TVT, etc., including 2 Highly Cited Papers of ESI (as the first author). His current research interests include modulation and signal processing for wireless communications, terahertz communications, near-space wireless communications, and visible light communications. He was a recipient of the 8th Young Elite Scientists Sponsorship Program by China Association for Science and Technology, the Science and Technology Award (Second Prize) of China Institute of Communications, the Excellent Master Dissertation of Chinese Institute of Electronics, the Special Scholarship of Tsinghua University, the Outstanding Ph.D. Graduate of Beijing City and the Outstanding Master Graduate of Tsinghua University. He is currently an Associate Editor of IEEE Communications Letters and IEEE Open Journal of Vehicular Technology, and a Guest Editor of IEEE Open Journal of the Communications Society. He was also the Exemplary Reviewer of IEEE Transactions on Communications and Communications Letters. 
\end{IEEEbiography}

\begin{IEEEbiography}[{\includegraphics[width=1in,height=1.25in,clip,keepaspectratio]{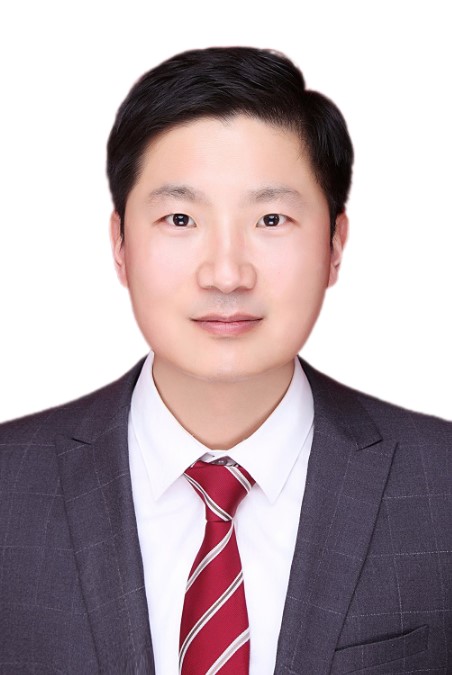}}]
{Di Zhang } (S'13, M'17, SM'19) received his Ph.D. degree (with honors) from Waseda University, Tokyo, Japan in 2017. He currently is an Associate Professor of Zhengzhou University, Zhengzhou 450001, China, he is also a Visiting Scholar of Korea University, Seoul 02841, Korea. He is serving as area editor of KSII Transactions on Internet and Information Systems. He has served as guest editor of \textsc{IEEE Wireless Communications} and \textsc{IEEE Network}, co-chair and TPC member of many IEEE flagship conferences. In 2023, He received the First Prize for Scientific and Technological Achievement Award of the Department of Education of Henan Province, and the First Prize for Science and Technology Progress Award of Henan Province. In 2019, he received the ITU Young Author Award. His is working on the wireless communications and networking, especially the short packet communications.
\end{IEEEbiography}

\begin{IEEEbiography}[{\includegraphics[width=1in,height=1.25in,clip,keepaspectratio]{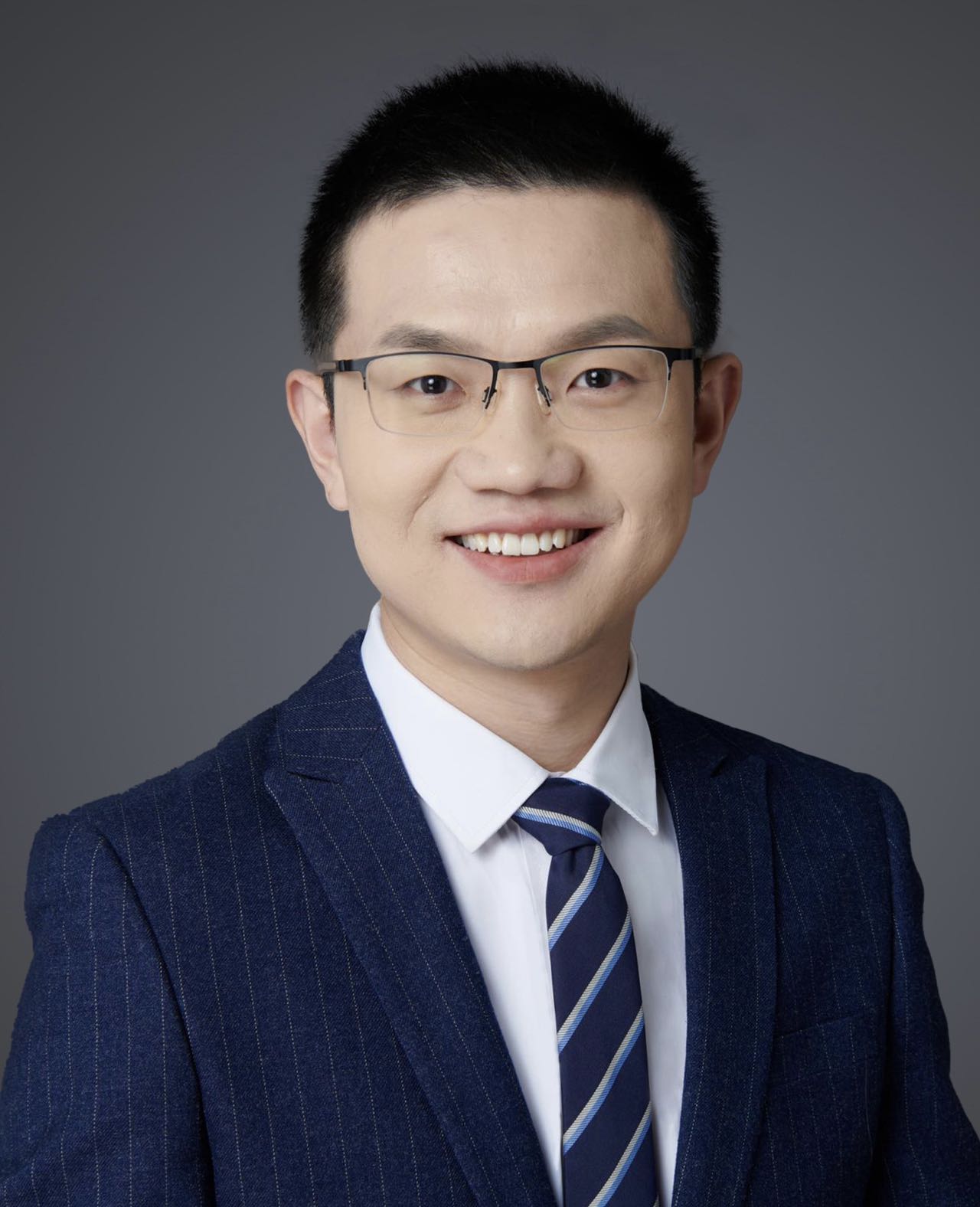}}]{Qingqing Wu} (S’13-M’16-SM’21)  is an Associate Professor with Shanghai Jiao Tong University. His current research interest includes intelligent reflecting surface (IRS), unmanned aerial vehicle (UAV) communications, and MIMO transceiver design. He has coauthored more than 100 IEEE journal papers with 29 ESI highly cited papers and 9 ESI hot papers, which have received more than 20,000 Google citations. He was listed as the Clarivate ESI Highly Cited Researcher in 2022 and 2021, the Most Influential Scholar Award in AI-2000 by Aminer in 2021 and World’s Top 2\% Scientist by Stanford University in 2020 and 2021.

He was the recipient of the IEEE Communications Society Fred Ellersick Prize, IEEE  Best Tutorial Paper Award in 2023, Asia-Pacific Best Young Researcher Award and Outstanding Paper Award in 2022, Young Author Best Paper Award in 2021, the Outstanding Ph.D. Thesis Award of China Institute of Communications in 2017, the IEEE ICCC Best Paper Award in 2021, and IEEE WCSP Best Paper Award in 2015. He was the Exemplary Editor of IEEE Communications Letters in 2019 and the Exemplary Reviewer of several IEEE journals. He serves as an Associate Editor for IEEE Transactions on Communications, IEEE Communications Letters, IEEE Wireless Communications Letters. He is the Lead Guest Editor for IEEE Journal on Selected Areas in Communications. He is the workshop co-chair for IEEE ICC 2019-2023 and IEEE GLOBECOM 2020. He serves as the Workshops and Symposia Officer of Reconfigurable Intelligent Surfaces Emerging Technology Initiative and Research Blog Officer of Aerial Communications Emerging Technology Initiative. He is the IEEE Communications Society Young Professional Chair in Asia Pacific Region.

\end{IEEEbiography}

\begin{IEEEbiography}[{\includegraphics[width=1in,height=1.25in,clip,keepaspectratio]{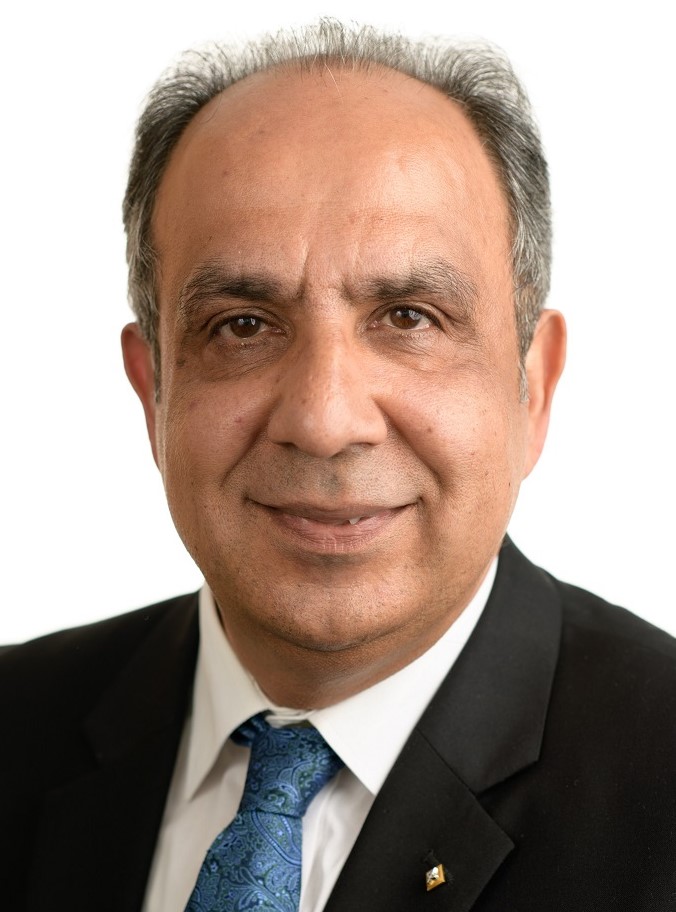}}]{Abbas Jamalipour} (S’86–M’91–SM’00–F’07) received the Ph.D. degree in Electrical Engineering from Nagoya University, Nagoya, Japan in 1996. He holds the positions of Professor of Ubiquitous Mobile Networking with the University of Sydney and since January 2022, the Editor-in-Chief of the IEEE Transactions on Vehicular Technology. He has authored nine technical books, eleven book chapters, over 550 technical papers, and five patents, all in the area of wireless communications and networking. Prof. Jamalipour is a recipient of the number of prestigious awards, such as the 2019 IEEE ComSoc Distinguished Technical Achievement Award in Green Communications, the 2016 IEEE ComSoc Distinguished Technical Achievement Award in Communications Switching and Routing, the 2010 IEEE ComSoc Harold Sobol Award, the 2006 IEEE ComSoc Best Tutorial Paper Award, as well as over 15 Best Paper Awards. He was the President of the IEEE Vehicular Technology Society (2020-2021). Previously, he held the positions of the Executive Vice-President and the Editor-in-Chief of VTS Mobile World and has been an elected member of the Board of Governors of the IEEE Vehicular Technology Society since 2014. He was the Editor-in-Chief IEEE WIRELESS COMMUNICATIONS, the Vice President-Conferences, and a member of Board of Governors of the IEEE Communications Society. He sits on the Editorial Board of the IEEE ACCESS and several other journals and is a member of Advisory Board of IEEE Internet of Things Journal. He has been the General Chair or Technical Program Chair for several prestigious conferences, including IEEE ICC, GLOBECOM, WCNC, and PIMRC. He is a Fellow of the Institute of Electrical and Electronics Engineers (IEEE), the Institute of Electrical, Information, and Communication Engineers (IEICE), and the Institution of Engineers Australia, an ACM Professional Member, and an IEEE Distinguished Speaker. 
\end{IEEEbiography}
\end{document}